\documentclass[aps,prc,onecolumn,floatfix,12pts,superscriptaddress,nofootinbib]{revtex4}
\usepackage{epsfig}
\usepackage{amssymb}
\usepackage{amsmath}
\usepackage{color}
\usepackage{graphicx}
\usepackage{epsfig}
\usepackage{amsmath}
\usepackage{color}
\usepackage{graphicx}
\usepackage{euscript}
\usepackage[toc,page]{appendix}
\usepackage{CJK}
\usepackage{float}
\usepackage{siunitx}
\definecolor{blue}{rgb}{0.05, 0.05, 0.5}
\def \beq{\begin{equation}}
\def \eeq{\end{equation}}
\def \beqa{\begin{eqnarray}}
\def \eeqa{\end{eqnarray}}

\usepackage{xspace}

\begin{document}
\begin{CJK*}{UTF8}{gbsn}

\title{Effect of global momentum conservation on longitudinal flow decorrelation}
\author{Pingal Dasgupta}
\affiliation{Key Laboratory of Nuclear Physics and Ion-beam Application (MOE), Institute of Modern Physics, Fudan University, Shanghai 200433, China}
\affiliation{Shanghai Research Center for Theoretical Nuclear Physics, NSFC and Fudan University, Shanghai $200438$, China}
\author{Han-Sheng Wang}
\affiliation{Key Laboratory of Nuclear Physics and Ion-beam Application (MOE), Institute of Modern Physics, Fudan University, Shanghai 200433, China}
\affiliation{Shanghai Research Center for Theoretical Nuclear Physics, NSFC and Fudan University, Shanghai $200438$, China}
\author{Guo-Liang Ma}
\email[]{glma@fudan.edu.cn}
\affiliation{Key Laboratory of Nuclear Physics and Ion-beam Application (MOE), Institute of Modern Physics, Fudan University, Shanghai 200433, China}
\affiliation{Shanghai Research Center for Theoretical Nuclear Physics, NSFC and Fudan University, Shanghai $200438$, China}

\begin{abstract}
We calculate the longitudinal flow decorrelation coefficients, i.e., $r_n(\eta,\eta_r)$ for $n=2,3$, in the presence of hydro-like flow and the global momentum conservation (GMC) constraint. The longitudinal flow decorrelation is weakened due to the GMC constraint. The GMC effect is sensitive to the total number of particles involved in GMC, the average longitudinal momentum, the transverse momentum, and the reference pseudorapidity. Our results of the $r_2(\eta,\eta_{rA})/r_2(\eta,\eta_{rB})$ ratio between two reference pseudorapidity bins are consistent with the experimental measurements. We predict that the modification effect of GMC on longitudinal flow decorrelation is more noticeable at BNL Relativistic Heavy Ion Collider energies than at CERN Large Hadron Collider energies.  Our finding provides a new perspective for understanding the longitudinal flow decorrelation in relativistic heavy-ion collisions. 
\end{abstract}
\maketitle  
\section{INTRODUCTION}
Experiments at the relativistic heavy-ion colliders create excellent opportunities to study the hot and dense deconfined phase of quarks and gluons, commonly known as the QGP (quark-gluon plasma)~\cite{Collins:1974ky,Shuryak:1980tp,Busza:2018rrf,Luo:2017faz}. The azimuthal anisotropic flow of hadrons in heavy ion collisions is critical evidence for collective expansion~\cite{Kolb:2000sd,Teaney:2000cw,Stoecker:2004qu,Lan:2022rrc}. The relativistic viscous hydrodynamical framework thus has emerged as one of the most successful descriptions of the fireball evolution~\cite{PHENIX:2003qra,STAR:2003wqp,STAR:2003xyj,ALICE:2010suc, ATLAS:2011ah,Ollitrault:1992bk, Rischke:1995ir,Bhalerao,Heinz:2013th,Gale:2013da,Huovinen:2013wma,Song:2017wtw,Shen:2020mgh}.  For the past two decades, the flow behavior has been studied extensively to understand the initial state of the collisions and the transport properties of the strongly interacting quantum chromodynamics (QCD) medium~\cite{Qin:2010pf,Holopainen:2010gz,Noronha-Hostler:2015dbi,Niemi:2012aj,Gardim:2011xv,Bzdak:2013zma,Alver:2010dn,Greco:2008fs,Jaiswal:2016hex,Yan:2017ivm,Xu:2014tda,Shi:2018izg}. 

The anisotropic flow parameters depict the harmonic modulation of the particle density distribution along the azimuthal direction, i.e., $dN/d\phi\propto1+2\sum_{n=1}^{\infty} v_n \cos[n(\phi-\psi_n)]$, where $v_n$  characterizes the strength of the $n$th order of anisotropic flow and $\psi_n$ is the corresponding event plane angle. The boost invariant hydrodynamical framework particularly has been very successful in predicting the behavior of flow observable (i.e., $v_n$) of different hadrons and their relative fluctuations at mid-rapidity for various symmetric collision systems ($A+A$) at the BNL Relativistic Heavy Ion Collider (RHIC) and the CERN Large Hadron Collider (LHC)\cite{Bernhard:2016tnd}.  However, recent developments in the theory and experiments have shown that we require a more realistic  $3+1$ hydrodynamic framework with fluctuating three-dimensional (3D) initial conditions to describe the copious production of particles and their flow observables along the rapidity direction~\cite{Schenke:2010nt,Schenke:2010rr,Karpenko:2013wva,Molnar:2014zha}. The experiments incorporating various asymmetric and small collisions at RHIC and the LHC have witnessed the breaking of flow factorization in both transverse momentum and rapidity directions~\cite{Bozek:2021mov}. One measured observable related to the breaking of flow factorization in the rapidity direction is called the longitudinal flow decorrelation coefficient [i.e., $r_n(\eta,\eta_r)$], which provides important insights regarding the longitudinal structure of the fireball produced in heavy-ion collisions, i.e. $v_n(\eta_1)\ne v_n(\eta_2)$ and $\psi_n(\eta_1)\ne \psi_n(\eta_2)$ for $\eta_1\ne\eta_{2}$~\cite{Xiao:2012uw,Jia:2014ysa,Jia:2017kdq}. For example, the CMS and  ATLAS collaborations have measured the longitudinal flow decorrelation coefficients for Pb+Pb and $p+ $Pb collisions for different centrality bins at $2.76A$ TeV and $5.02A$ TeV~\cite{CMS:2015xmx,ATLAS:2017rij}. The  ATLAS collaboration has also studied the system size dependence of the longitudinal flow decorrelation coefficients with Xe+Xe collisions at $5.44A$ TeV~\cite{ATLAS:2020sgl}. The  STAR collaboration has taken a similar initiative to study the coefficients for Au+Au collisions at different beam energies~\cite{Nie:2020trj}, for Ru+Ru and Zr+Zr collisions at $0.2A$ TeV~\cite{Nie:2022gbg}. The physics explanation for the longitudinal flow decorrelation is thought to be due to the twist of event plane angles of initial torqued fireball~\cite{Bozek:2010vz,Bozek:2017qir} or initial-state longitudinal fluctuations~\cite{Dumitru:2011vk,Schenke:2016ksl,Sakai:2021rug}.  The initial spatial decorrelation in the longitudinal direction is finally transferred into the measured longitudinal flow decorrelation through the final state evolution~\cite{Pang:2014pxa,Bozek:2015bha,Wu:2021hkv,Behera:2020mol,Sakai:2020pjw}.

The longitudinal flow decorrelation coefficient measures the correlation between two flow vectors at two symmetric rapidity bins. The absence of collectivity in an expanding system produces maximum decorrelation, i.e. $r_n(\eta,\eta_r)\approx0$, whereas the boost invariant type of longitudinal flow invokes maximum correlation (or minimum decorrelation). The boost invariance is not a realistic way to deal with 3D expansion, since the initial state fluctuations along the rapidity play an essential role in the outcome of the flow coefficients. In contrast to flow, there is a correlation known as the ``non-flow" contribution. The momentum conservation can be considered as one of the sources for ``non-flow'' ~\cite{Borghini:2000sa,Borghini:2003ur,Borghini:2007ku}, which is satisfied by any final state interactions. Considering the transverse momentum conservation (TMC) and hydro-like flow, the multi--particle cumulants at mid--rapidity have been studied elaborately in  Refs.~\cite{Bzdak:2017zok,Bzdak:2018web,Xie:2022kwu}. The studies found that the non--flow contribution from TMC can not be ignored to understand the true flow contribution in small collision systems, because the TMC contribution is more significant for particles with higher transverse momenta and for systems with a smaller number of particles. In this paper, we aim to explore the contribution of global momentum conservation (GMC) on the longitudinal flow decorrelation in relativistic heavy-ion collisions.

Our paper is organized as follows. In Section II, we derive the two-particle azimuthal correlation under the influence of GMC. Section III is dedicated to finding out the longitudinal flow decorrelation parameter $r_n(\eta,\eta_r)$ in the presence of hydro-like flow and GMC.  Section IV discusses a parametric form of particle production from 3D ideal hydrodynamic evolution to obtain some key parameters. We present our main results about the effects of global momentum conservation on the longitudinal flow decorrelation in Section V. Finally, we summarize in Section VI.

\section{TWO-particle azimuthal cumulant FROM GMC }

For a system consisting of $N$ number of particles, the  $N$-particle joint probability distribution function upon full phase-space integration normalizes to unity. Any $M$-particle observable ($M \textless N$) is calculated using  $M$-particle joint probability distribution function in momentum space, which is expressed as $f({\bf p_1},{\bf p_2},{\bf p_3},...,{\bf p_M})$~\cite{Borghini:2000sa,Borghini:2007ku,Borghini:2003ur}. The  $M$-particle joint probability distribution can be expressed as the product of $M$-single particle distributions if the $M$ momenta are independent. However, the $M$ momenta are not independent of each other. Consequently, the joint probability distribution function includes additional terms. For two- and three-particle cases, the joint probability distributions look as follows :

\begin{align}
\centering
f({\bf p_1},{\bf p_2})&=f_c({\bf p_1}) f_c({\bf p_2})+f_c({\bf p_1},{\bf p_2}) \nonumber\\
f({\bf p_1},{\bf p_2},{\bf p_3})&=f_c({\bf p_1})f_c({\bf p_2}) f_c({\bf p_3})+f_c({\bf p_3}) f_c({\bf p_1},{\bf p_2})+f_c({\bf p_2}) f_c({\bf p_1},{\bf p_3}) \nonumber \\
&\  \  + f_c({\bf p_1}) f_c({\bf p_2},{\bf p_3})+f_c({\bf p_1},{\bf p_2},{\bf p_3}) \ \ ,
\label{fulldist}
\end{align}
where each term is a cumulant corresponding to the distinct partition of $M$-particles. Here, the single particle cumulant corresponds to the single particle probability distribution function, i.e., $f_c({\bf p})=f({\bf p})$ and the product of single-particle cumulant is known as the ``indirect" correlation term. In contrast, the $M$-particle cumulant [i.e., $f_c(p_1,p_2,..,p_M)$] represents the ``direct"  correlation term of $M$-particles.

In Ref~\cite{Borghini:2007ku,Borghini:2003ur}, a systematic way of obtaining the cumulants using generating function has been presented. First, we define the generating function of joint probability distribution as follows:
\begin{eqnarray}
G(x_1,x_2, . \ . \ . ,x_N)=1+x_1 f({\bf p_1})+x_2 f({\bf p_2})+ . \ . \ . +x_1 x_2 f({\bf p_1},{\bf p_2})+. \ .  \ .,
\label{gen}
\end{eqnarray}
where, $x_1,x_2,. \ . \ . , x_N$ are auxiliary variables.  To obtain the ``direct" cumulant terms, we take the logarithm of Eq.~(\ref{gen}):
\begin{eqnarray}
\ln G(x_1,x_2, . \ . \ . ,x_N)=1+x_1 f_c({\bf p_1})+x_2 f_c({\bf p_2})+ .\ . \ . +x_1 x_2 f_c({\bf p_1},{\bf p_2})+.\ .\ .,
\label{gen2}
\end{eqnarray}
where the coefficient of  $x_1x_2. \ . \ . x_j$  represents the  corresponding cumulant term $f_c({\bf p_1},{\bf p_2},{\bf p_3}, .\ .\ ., {\bf p_j})$.
\subsection{ Two-particle cumulant from GMC}
Momentum conservation is one such constraint that restricts the momentum of particles to be not independent of each other. Next, we discuss how momentum conservation changes the joint probability distribution function.  In the centre of mass frame for a $N$ particle system,  the sum of $N$-momenta is always zero, i.e., ${\bf p_1}+{\bf p_2}+{\bf p_3}+....+{\bf p_N}=0$ in 3-dimensional momentum space. Thus, the joint $N$-particle probability distribution can be represented as ~\cite{Borghini:2007ku,Borghini:2003ur} :
\begin{eqnarray}
f({\bf p_1},{\bf p_2},{\bf p_3},.\ .\ .,{\bf p_N})=\frac{1}{A}\delta^3({\bf p_1}+{\bf p_2}+{\bf p_3}+.\ .\ .+{\bf p_N}) \  F({\bf p_1}) F({\bf p_2}) F({\bf p_3}).\ .\ .F({\bf p_N}) ,
\end{eqnarray}
where $A$ is an overall normalization constant, $F({\bf p})$ is the single-particle `unnormalized' probability distribution function, and $\delta^3({\bf p_1}+{\bf p_2}+{\bf p_3}+.\ .\ .+{\bf p_N})$ is for the global momentum conservation (GMC) constraint\footnote{If we only consider of transverse momentum conservation,  $\bf p$ will be reduced to $\bf p_T$.}. Now the two-particle joint distribution from the above equation can be obtained by integrating over all $N-2$ momenta as :

\begin{align}
f({\bf p_1},{\bf p_2})=\frac{1}{A}\bigg(\displaystyle \prod_{j=1}^{2} F({\bf p_j})\bigg) \times  \int \delta^3({\bf p_1}+{\bf p_2}+{\bf p_3}+....+{\bf p_N}) \displaystyle \prod_{j=3}^{N} \bigg[F({\bf p_j}) d{\bf p_j}\bigg].
 \label{gmc}
\end{align}
Note that in the absence of  GMC constraint, $F({\bf p})= f({\bf p})$ and the joint probability distribution function simply factorizes to the product of two single-particle distribution functions.

For large $N$, we approximate the two-particle joint probability distribution (or two-particle cumulant) ~\cite{Borghini:2007ku} as:
\begin{equation}
{f}({\bf p_1},{\bf p_2})=F({\bf p_1}) F({\bf p_2}) \ \exp \bigg[ -\frac{p_{1,x} p_{2,x}}{N\langle p_x^2\rangle_F}- \frac{p_{1,y} p_{2,y}}{N\langle p_y^2\rangle_F} -\frac{p_{1,z} p_{2,z}}{N\langle p_z^2\rangle_F} \bigg] ,
\label{gmc1}
\end{equation}
where, $x$-,$y$- and $z$- axes are three principal axes that diagonalize the tensor $\langle {\rm \bf p} \otimes {\rm \bf p}\rangle$. The $z$-axis, conventionally, denotes the beam axis, whereas the $x$-axis represents the direction of the impact parameter and the $y$-axis lies perpendicular to the $x$-axis and $z$-axis. 
The quantity $\langle...\rangle_F$ in the above equation denotes the average obtained from full phase space integration as :
\begin{align}
\langle O({\rm \bf p}) \rangle_F=\frac{\displaystyle \int_F O({\rm \bf p}) F({\rm \bf p})d^3{\rm \bf p}}{{\displaystyle \int_F  F({\rm \bf p})d^3{\rm \bf p}}}.
\end{align}
In the Cartesian coordinate system, the three-momentum ($\bf{p}$)  of  a particle can be written as :
\begin{equation}
\centering
 {\bf p}=
 \begin{pmatrix}
 p_x= p_T\cos\phi \\ p_y=p_T\sin\phi \\ p_z=\sqrt{p_T^2+m^2}\sinh y = p_T \sinh \eta \hfill
\end{pmatrix},
\end{equation}
where $m$, $p_T$, $y$ ($=\frac{1}{2} {\rm ln}\frac{E+p_z}{E-p_z}$), $\eta$ ($=\frac{1}{2} {\rm ln}\frac{p+p_z}{p-p_z}$) and $\phi$ are the mass, transverse momentum, rapidity, pseudorapidity, and momentum azimuthal angle of the particle, respectively.

The $n$th harmonic of two-particle azimuthal cumulant, i.e., $\rm c_n\{2\}=\langle e^{in(\phi_1-\phi_2)}\rangle$,  can be obtained by performing the azimuthal integration of the two-particle joint probability distribution as : 
\begin{align}
c_n\{2\}|_{\eta_1,p_{1};\eta_2,p_{2}}=\frac{\displaystyle \int_0^{2\pi} f({\bf p_1},{\bf p_2})e^{in(\phi_1-\phi_2)} d\phi_1 d\phi_2}{\displaystyle \int_0^{2\pi} f({\bf p_1},{\bf p_2}) d\phi_1 d\phi_2}.
\label{2particle}
\end{align}
\subsection{Two-particle azimuthal cumulant from  GMC + hydro-like flow}
For hydro-like flow, we can approximate  the single-particle distribution at the pseudorapidity $\eta$ as :
\begin{align}
F({\bf p})=\frac{g(p,\eta)}{2\pi}\bigg[1+\sum_{n}^{}2v_n(p,\eta)  \ {\rm cos}[n(\phi-\psi_n(\eta))]\bigg],
\label{hydrolike}
\end{align}
where the $v_n(p,\eta)$ and $\psi_n(\eta)$ are the $n$-th order the differential flow parameter and the $n$-th event-plane angle at pseudorapidity $\eta$, respectively. Note that we henceforth denote $p_T$ as $p$ for simplicity. Next, we will calculate the second and third order harmonics of the two-particle cumulants in the presence of hydro-like flow and GMC using Eqs.~(\ref{hydrolike}) and~(\ref{gmc1}).

Considering elliptic flow only and expanding Eq.~(\ref{gmc1}) up to the second order ($1/N^2$), we get the following result of two-particle $c_2\{2\}$ \footnote{Note that evaluating Eq.~(\ref{2particle}), the denominator is considered as $4\pi^2$ only, as the other terms are suppressed by the higher power of $1/N$.} : 
\begin{align}
c_2\{2\}& =\langle e^{i2(\phi_1-\phi_2)}\rangle|^{\rm \sf GMC+Flow}_{\eta_1,p_{1};\eta_2,p_{2}} \nonumber\\
&\approx \ v_{2}\left(p_{1}, \eta_{1}\right)  \ v_{2}\left(p_{2}, \eta_{2}\right) \cos\left[2(\psi_{2}\left(\eta_{1}\right)-\psi_{2}\left(\eta_{2})\right)\right] \nonumber\\
&\  \  \ -\frac{p_{1} \ \sinh \left(\eta_{1}\right) \  p_{2} \ \sinh \left(\eta_{2}\right) \  v_{2}\left(p_{1}, \eta_{1}\right) \ v_{2}\left(p_{2}, \eta_{2}\right)
	\cos \left[2 (\psi_{2}\left(\eta_{1}\right)- \psi_{2}\left(\eta_{2})\right)\right]}
{N\left\langle p_{z}^{2}\right\rangle_{F}}\nonumber \\
&\  \  \ +\frac{p_{1}^{2}  \ \sinh ^{2} \left(\eta_{1}\right)  \ p_{2}^{2} \  \sinh ^{2} \left(\eta_{2}\right) \  v_{2}\left(p_{1} \eta_{1}\right) \ v_{2}\left(p_{2}, \eta_{2}\right)
		\cos \left[2(\psi_{2}\left(\eta_{1}\right)- \psi_{2}\left(\eta_{2})\right)\right]}
	{2 N^{2}\left\langle p_{z}^{2}\right\rangle_{F}^{2}}\nonumber\\
&\  \  \  +\frac{p_{1}^{2} \ p_{2}^{2}  \ v_{2}\left(p_{1}, \eta_{1}\right)  \ v_{2}\left(p_{2}, \eta_{2}\right) \operatorname{cos}\left[2 (\psi_{2}\left(\eta_{1}\right)-2 \psi_{2}\left(\eta_{2})\right)\right]}{N^{2}\left\langle p_T^{2}\right\rangle_F^{2}}  +\frac{p_{1}^{2}  \ p_{2}^{2}}{2 N^{2}\left\langle p_{T}^{2}\right\rangle_{F}^{2}}
	\label{nparticlegmc}
\end{align}
where we assume that $\langle p_{x}^{2}\rangle_{F}\approx\langle p_{y}^{2}\rangle_{F}\approx\langle p_{T}^{2}\rangle_{F}/2$.
Similarly, after considering the momentum anisotropy flow up to third order harmonic and expanding the  Eq.~(\ref{gmc1}) up to the third power $1/N^3$, we can obtain the following result of two-particle $c_3\{2\}$ :
\begin{align}
c_3\{2\} & = \langle e^{i3(\phi_1-\phi_2)}\rangle|^{\rm \sf GMC+Flow}_{\eta_1,p_{1};\eta_2,p_{2}} \nonumber\\
&\approx v_{3}\left(p_{1}, \eta_{1}\right) \ v_{3}\left(p_{2}, \eta_{2}\right) \cos \left[3\left(\psi_{3}\left(\eta_{1}\right)-\psi_{3}\left(\eta_{2}\right)\right)\right]\nonumber\\
&\  \  \ -\frac{p_{1}  \ p_{2} \ v_{2}\left(p_{1}, \eta_{1}\right) \ v_{2}\left(p_{2}, \eta_{2}\right) \ \cos \left[2\left(\psi_{2}\left(\eta_{1}\right)-\psi_{2}\left(\eta_{2}\right)\right)\right]}{N\left\langle p_{T}^{2}\right\rangle_{F}}\nonumber\\
&\  \  \ -\frac{p_{1} \ \sinh \left(\eta_{1}\right) \  p_{2} \ \sinh \left(\eta_{2}\right) \ v_{3}\left(p_{1}, \eta_{1}\right)\ v_{3}\left(p_{2}, \eta_{2}\right) \cos \left[3\left(\psi_{3}\left(\eta_{1}\right)-\psi_{3}\left(\eta_{2}\right)\right)\right]}{N\left\langle p_{z}^{2}\right\rangle_{F}}\nonumber\\
&\  \  \ +\frac{p_{1}^{2} \  p_{2}^{2}\  v_{3}\left(p_{1}, \eta_{1}\right) \ v_{3}\left(p_{2}, \eta_{2}\right)\  \cos \left[3\left(\psi_{3}\left(\eta_{1}\right)-\psi_{3}\left(\eta_{2}\right)\right)\right]}{N^{2}\left\langle p_{T}^{2}\right\rangle_{F}^2}\nonumber\\
&\  \  \ +\frac{p_{1}^{2} \  \sinh \left(\eta_{1}\right) \ p_{2}^{2}   \ \sinh \left(\eta_{2}\right) \  v_{2}\left(p_{1}, \eta_{1}\right) \ v_{2}\left(p_{2}, \eta_{2}\right) \cos\left[2\left(\psi_{2}\left(\eta_{1}\right)-\psi_{2}\left(\eta_{2}\right)\right)\right]}{N^{2}\left\langle p_{T}^{2}\right\rangle_{F}\left\langle p_{z}^{2}\right\rangle_{F}}\nonumber\\
&\  \  \ +\frac{p_{1}^{2} \ \sinh^{2}\left(\eta_{1}\right) \ p_{2}^{2} \ \sinh ^{2} \left(\eta_{2}\right)\  v_{3}\left(p_{1}, \eta_{1}\right) \ v_{3}\left(p_{2}, \eta_{2}\right) \cos\left[3\left(\psi_{3}\left(\eta_{1}\right)-\psi_{3}\left(\eta_{2}\right)\right)\right]}{2 N^{2}\left\langle p_{z}^{2}\right\rangle_{F}^2}\nonumber\\
&\  \  \ -\frac{p_{1}^{3} \ p_{2}^{3}  \ v_{2}\left(p_{1}, \eta_{1}\right)  \ v_{2}\left(p_{2}, \eta_{2}\right) \ \cos \left[2\left(\psi_{2}\left(\eta_{1}\right)-\psi_{2}\left(\eta_{2}\right)\right)\right]}{2 N^{3}\left\langle p_{T}^{2}\right\rangle_{F}^{3}}\nonumber\\
&\  \  \ -\frac{p_{1}^{3}\  \sinh \left(\eta_{1}\right)  \ p_{2}^{3} \ \sinh \left(\eta_{2}\right) \ v_{3}\left(p_{1}, \eta_{1}\right)\  v_{3}\left(p_{2}, \eta_{2}\right) \ \cos \left[3\left(\psi_{3}\left(\eta_{1}\right)-\psi_{3}\left(\eta_{2}\right)\right)\right]}{N^{3}\left\langle p_{T}^{2}\right\rangle_{F}^{2}\left\langle p_{z}^{2}\right\rangle_{F}}\nonumber\\
&\  \  \ -\frac{p_{1}^{3} \ \sinh^{2}(\eta_1) \  p_{2}^{3} \  \sinh^{2}(\eta_2) \  v_{2}\left(p_{1}, \eta_{1}\right) \ v_{2}\left(p_{2}, \eta_{2}\right) \cos \left[2\left(\psi_{2}\left(\eta_{1}\right)-\psi_{2}\left(\eta_{2}\right)\right)\right]}{2 N^{3}\left\langle p_{T}^{2}\right\rangle_{F}\left\langle p_{z}^{2}\right\rangle_{F}^{2}}\nonumber\\
&\  \  \ -\frac{p_{1}^{3} \ \sinh^3\left(\eta_{1}\right) \ p_{2}^{3} \ \sinh ^{3}\left(\eta_{2}\right) \ v_{3}\left(p_{1}, \eta_{1}\right) \ v_{3}\left(p_{2}, \eta_{2}\right )\cos \left[3\left(\psi_{3}\left(\eta_{1}\right)-\psi_{3}\left(\eta_{2}\right)\right)\right]}{6N^3 \langle p_z^2\rangle_F^3}\nonumber\\
&\  \  \ -\frac{p_{1}^{3} \  p_{2}^{3}}{6 N^{3}\left\langle p_{T}^{2}\right\rangle_{F}^{3}}.
\label{nparticlegmc2}
\end{align}
In both Eqs.~(\ref{nparticlegmc}) and ~(\ref{nparticlegmc2}), we can see a pure flow-contributed term [$v_n(p_1,\eta_1) \ v_n(p_2,\eta_2) \ \cos[n(\psi_n(\eta_1)-\psi_n(\eta_2))]$] in the beginning and a pure TMC term [$(-1)^n\frac{p_1^n p_2^n}{n! N^n \langle p_{T}^{2}\rangle_{F}^{n}}$] which is flow-independent at the end~\cite{Bzdak:2017zok}. The former arises from the ``indirect" part of the joint probability distribution, whereas the latter comes from the ``direct" part. The remaining terms are of the order of $\mathcal{O}(1/N)$, $\mathcal{O}(1/N^2)$, or higher powers, which appear due to the interplay between GMC and hydro-like flow.

To estimate the first order correction in the two-particle correlation under the influence of GMC, we drop all the terms with $\mathcal{O}(1/N^2)$ and higher powers\footnote{We have checked that these higher orders of terms do not significantly change our results below.}. Consequently, Eqs.~(\ref{nparticlegmc}) and ~(\ref{nparticlegmc2}) look simpler as follows:
\begin{align}
c_2\{2\}\approx
& \  v_{2}\left(p_{1}, \eta_{1}\right)  \ v_{2}\left(p_{2}, \eta_{2}\right) \cos\left[2 (\psi_{2}\left(\eta_{1}\right)-\psi_{2}\left(\eta_{2})\right)\right] \nonumber\\
&\times(1-\frac{p_{1} \ \sinh \left(\eta_{1}\right) \  p_{2} \  \sinh \left(\eta_{2}\right) \  v_{2}\left(p_{1}, \eta_{1}\right) \ v_{2}\left(p_{2}, \eta_{2}\right)
	\cos \left[2 (\psi_{2}\left(\eta_{1}\right)- \psi_{2}\left(\eta_{2})\right)\right]}
{N\left\langle p_{z}^{2}\right\rangle_{F}}),
\label{nparticlegmc-3}
\end{align}
\begin{align}
c_3\{2\}\approx
& \  v_{3}\left(p_{1}, \eta_{1}\right) \ v_{3}\left(p_{2}, \eta_{2}\right) \cos \left[3\left(\psi_{3}\left(\eta_{1}\right)-\psi_{3}\left(\eta_{2}\right)\right)\right]\nonumber\\
&-\frac{p_{1}  \ p_{2}  \ v_{2}\left(p_{1}, \eta_{1}\right) \ v_{2}\left(p_{2}, \eta_{2}\right) \ \cos \left[2\left(\psi_{2}\left(\eta_{1}\right)-\psi_{2}\left(\eta_{2}\right)\right)\right]}{N\left\langle p_{T}^{2}\right\rangle_{F}}\nonumber\\
&-\frac{p_{1} \ \sinh \left(\eta_{1}\right) \  p_{2} \ \sinh \left(\eta_{2}\right) \ v_{3}\left(p_{1}, \eta_{1}\right)\ v_{3}\left(p_{2}, \eta_{2}\right) \cos \left[3\left(\psi_{3}\left(\eta_{1}\right)-\psi_{3}\left(\eta_{2}\right)\right)\right]}{N\left\langle p_{z}^{2}\right\rangle_{F}}.
\label{nparticlegmc-4}
\end{align}
From the above two equations, we can see that the two-particle azimuthal cumulants can be modified by the constraint of GMC. 
For example, the $c_2\{2\}$ can be decomposed into the product of two terms as shown in Eq.~(\ref{nparticlegmc-3}), where the first-line term is the regular $v_2$ decorrelation and the second-line one can be seen as a modification factor from GMC. The modification factor is smaller than one when the two particles are in the same longitudinal momentum direction, but greater than one when the two particles are in opposite longitudinal momentum directions. Considering that two-particle correlations are usually measured within a mid-rapidity window ($|\eta|<\eta_{cut}$), we expect that the total modification effect of GMC on two-particle correlations can be largely canceled out. However, the details of the $\Delta\eta$ dependence of the modification effect need further investigation. In this work, we will focus on how the GMC effect affects the longitudinal flow decorrelation.

\section{LONGITUDINAL FLOW DECORRELATION FROM GMC}
\subsection{Definition of longitudinal flow decorrelation $r_n$} 
The longitudinal decorrelation of harmonic flow measures the decorrelation effect between two symmetric pseudorapidity bins ($-\eta$ and $\eta$) along the longitudinal direction by comparing the correlations between each of them and a reference pseudorapidity bin $\eta_r$ (which is usually chosen a considerable value to avoid non-flow contribution). The observable is defined as :
\begin{align}
r_n(\eta,\eta_r)=\frac{ \langle {\bf Q_n}(-\eta)  {\bf Q^*_n}(\eta_r)\rangle}{\langle {\bf Q_n}(\eta)  {\bf Q^*_n}(\eta_r)\rangle},
\label{rn_def}
\end{align}
where the ${\bf Q_n}(\eta)$  vector  quantifies the $n$--th order harmonic flow in a collision event, and $\langle .\ . \ .\rangle$ denotes the event average.
\begin{align}
{\bf Q_n}(\eta)=\frac{1}{{M}}\displaystyle\sum_{i=1}^{M}e^{in\phi_i},
\end{align}
where $M$ represents the number of particles in the corresponding pseudorapidity $\eta$ bin. The longitudinal flow decorrelation coefficient $r_n$  can further be expressed as the ratio of two-particle correlations:
\begin{align}
r_n(\eta,\eta_r)=\frac{\langle e^{in(\phi_1(-\eta)-\phi_2(\eta_r))}\rangle}{\langle e^{in(\phi_1(\eta)-\phi_2(\eta_r))}\rangle}.
\label{rndef}
\end{align}
If there are only ``indirect" correlation terms but no ``direct" correlation terms present in the two-particle correlation in Eq.~(\ref{fulldist}), $r_n$ can be expressed as:
\begin{align}
&r_n(\eta,\eta_r)|^{\rm \sf Flow}_{-\eta;\eta;\eta_r} =\frac{\langle{v_{n}(-\eta)}  \ {v_{n}(\eta_r)}  \  \cos (n ({\psi_{n}(-\eta)}-{\psi_{n} (\eta_r)}))\rangle}{{\langle v_{n}(\eta)}  \ {v_{n}(\eta_r)}  \  \cos (n ({\psi_{n}(\eta)}-{\psi_{n} (\eta_r)}))\rangle},
\label{rnwogmc}
\end{align}
which reflects a normal form of longitudinal flow decorrelation due to hydro-like flow.

\subsection{Longitudinal flow decorrelation $r_n$ from  GMC + hydro-like  flow}
In the presence of both ``indirect" and ``direct" correlations like both hydro-like flow and GMC, the $r_n$ for $n=2,3$ can be evaluated by inserting the Eqs.~(\ref{nparticlegmc-3}) and~(\ref{nparticlegmc-4}) into the following relation :
\begin{align}
&r_n(\eta,\eta_r)|^{\rm \sf GMC+Flow}_{-\eta,p_{1};\eta,p_{2};\eta_r,p_{3}}=\frac{\langle e^{in(\phi_1-\phi_2)}\rangle|^{\rm \sf GMC+Flow}_{-\eta,p_{1};\eta_r,p_{3}}}{\langle e^{in(\phi_1-\phi_2)}\rangle|^{\rm \sf GMC+Flow}_{\eta,p_{2};\eta_r,p_{3}}}.
\label{r2gmc}
\end{align}
The above equation expresses a momentum-dependent (or differential) form of $r_n$. However, such a variable can mimic the actual momentum-integrated observable (for example, the transverse momentum window is considered to be $\rm 0.3  \ GeV$$ < p < 3$ GeV for the Pb+Pb collisions at the LHC experiment), if a relevant mean $p$  is chosen. We consider the transverse momenta of the particles at the forward, backward  (i.e., $\pm \ \eta$ ), and at the reference pseudorapidity ($\eta_r$)  to be equal to each other (i.e., $p_1\approx p_2 \approx p_3=p$, for $i=1,2$ and $3$) and subsequently by inserting Eq.~(\ref{nparticlegmc-3}) into Eq.~(\ref{r2gmc}), we can obtain the second order longitudinal flow decorrelation coefficient [i.e., $r_2(\eta,\eta_r)$]: \\
\begin{align}
r_2(\eta,\eta_r)|^{\rm \sf GMC+Flow}_{-\eta,p;\eta,p;\eta_r,p}&=r_2(\eta,\eta_r)|^{\rm \sf Flow}_{-\eta,p;\eta,p;\eta_r,p}    \  \ \times  \ \ R_2 \nonumber\\
{\rm where, \ \ \ \ \ }R_2&\approx\bigg[\frac{{p^2 \sinh(-\eta)} \  { \sinh(\eta_r)}-N {\langle p_z^2\rangle_F}}{{p^2 \sinh(\eta)} \  { \sinh(\eta_r)}-N {\langle p_z^2\rangle_F}}\bigg].
\label{rnwogmc3}
\end{align}
 As we can see from the above equation,  the modification factor for $r_2$ (denoted by $R_2$ here) appears due to the GMC effect, which depends on the transverse momentum $p$, (reference) pseudorapidity of particles, the averaged longitudinal momentum $\langle p_z^2\rangle_F$, and the total number of particles in the collision system. If the value of $R_2$ is one, it indicates no impact on the longitudinal decorrelation parameter. If $R_2$ is greater than one, it suggests that the longitudinal decorrelation is weakened due to the presence of GMC. However, a fractional value of $R_2$ indicates a stronger decorrelation in the presence of GMC. 
 
Similarly, an approximated expression for longitudinal decorrelation parameter $r_3$ using Eq.~(\ref{nparticlegmc-4}) appears as:
\begin{align}
\label{rnwogmc4}
 & \ \ \ \ \ \ \ \ \ \ \ r_3(\eta,\eta_r)|^{\rm \sf GMC+Flow}_{-\eta,p;\eta,p;\eta_r,p}=r_3(\eta,\eta_r)|^{\rm \sf Flow}_{-\eta,p;\eta,p;\eta_r,p}    \  \ \times  \ \ R_3  \nonumber\\
 &{\rm where, \ \ \ \ \ } \nonumber \\
& R_3\approx\frac{D}{C}\bigg[\frac{{p^2} \  {\langle p_z^2\rangle_F} \ {v_2(p,-\eta)} \ {v_2(p,\eta_r)} {v_3(p,\eta)} \  A -{\langle p_T^2\rangle_F} \ {v_3(p,\eta)} \ {v_3(p,-\eta)} \ {v_3(p,\eta_r)} \ C \left(N {\langle p_z^2\rangle_F} + p^2 \ \sinh ({\eta}) \  \sinh ({\eta_r})\right)}{{p^2} \  {\langle p_z^2\rangle_F} \ {v_2(p,\eta)} \ {v_2(p,\eta_r)} \ {v_3(p,-\eta)}  \ B-{\langle p_T^2\rangle_F} \ {v_3(p,-\eta)} \   {v_3(p,\eta)}\  {v_3(p,\eta_r)} \ D  \left(N {\langle p_z^2\rangle_F}-p^2 \sinh ({\eta})\  \sinh ({\eta_r})\right)}\bigg]\\
&{\rm where,}\nonumber\\
&A= \cos[2(\psi_2(-\eta)-\psi_2(\eta_r))], B= \cos[2(\psi_2(\eta)-\psi_2(\eta_r))],\  C=  \cos[3(\psi_3(-\eta)-\psi_3(\eta_r))]{,\rm and} \ D=  \cos[3(\psi_3(\eta)-\psi_3(\eta_r))]  \nonumber
\end{align}
In the next section, we will discuss how to obtain the estimates of $\langle p_z^2\rangle_F$ and  $N$ from a longitudinally accelerating perfect fluid system to calculate the effects of GMC on the longitudinal flow decorrelation coefficients.
\section{ ESTIMATE OF KEY PARAMETERS ($\langle p_z^2\rangle_\Gamma$ and  $N_\Gamma$) FROM IDEAL HYDRODYNAMICS}
The production and anisotropic flow of charged hadrons have been successfully described in a relativistic hydrodynamic framework. Both viscous and ideal hydrodynamic models have been found to provide satisfactory descriptions of the hadronic yields for various collision systems at the energies available at the RHIC and LHC. For the present study, we aim to find the estimates of $\langle p_z^2\rangle_F$ and  $N$ from a longitudinally expanding fireball. We follow Ref.~\cite{Ze-Fang:2017ppe} to obtain the pseudorapidity distribution of charged hadrons in a longitudinally accelerating perfect fluid system in which the net baryon number and energy-momentum tensor follow the conservation laws.  An approximate parametric relation for the rapidity distribution of charged hadrons in a perfect fluid system is as follows:
\begin{align}
\label{ydist}
\centering
\frac{dN_{\rm ch}}{dy}\approx N_0 \  {\rm cosh}^{-\frac{1}{2}\alpha(\lambda)-1}&\bigg(\frac{y}{\alpha(\lambda)}\bigg) \ {\rm exp}\bigg(-\frac{m}{T_{\rm f}}  {\rm cosh}^{\alpha(\lambda)}\bigg(\frac{y}{\alpha(\lambda)}\bigg)\bigg),
\end{align}

where $\alpha(\lambda)=\frac{2\lambda-1}{\lambda-1}$, and $N_0, T_{\rm f}$, and $ \lambda$ are three fit parameters.
To obtain the pseudorapidity distribution of charged hadrons with average mass $\bar{m}$, we use the  relation :
\begin{align}
\frac{dN_{\rm ch}}{d\eta}&\approx \frac{{\rm cosh \eta}}{\sqrt{D^2+ {\rm cosh^2 \eta}}} \frac{dN_{\rm ch}}{dy}\bigg|_{y=y(\eta)} ,
\end{align}
where the parameter $D=\bar{m}/\bigg[\frac{T_{\rm eff}}{1+\frac{\sigma^2}{2}(y-y_{\rm mid})^2}\bigg]$ determines the dip of the pseudorapidity distribution at  mid-rapidity. The other two fit parameters are $T_{\rm eff}$ and $\sigma$. The significance of these used parameters are listed as follows :
\begin{itemize}
	\item  $N_0$ is a normalization constant that fixes the particle density at mid-rapidity.
	\item $T_{\rm f}$ corresponds to the freeze-out temperature.
	\item $\lambda$ determines the longitudinal acceleration of the fluid.
	\item The effective temperature $T_{\rm eff}$ corresponds to the inverse slope parameter of $m_T-m$ spectra at the mid--rapidity.
	\item $\sigma$ parametrizes the effective temperature gradient.
\end{itemize}
The pseudorapidity distribution of all particles  (i.e., charged+neutral $=\frac{dN}{d\eta}$) at any pseudorapidity bin $\eta$ can be further approximated as, $\frac{dN}{d\eta}\approx1.5 \frac{dN_{\rm ch}}{d\eta}$. 
Finally,  the average of the variable $p_z^2$ for all particles, $\langle {p_z^2} \rangle_\Gamma$, over a phase space region $\Gamma\in \{-\eta,\eta\}$ (in terms of pseudorapidity range) is obtained by performing the following integral:
\begin{align}
\langle p_z^2\rangle_\Gamma=\frac {\displaystyle \int_{\Gamma} p_z^2 \frac{dN}{d\eta} d\eta}{\displaystyle\int_{\Gamma} \frac{dN}{d\eta} d\eta }.
\label{average}
\end{align}
The total number of particles $N$ within the phase-space region  $\Gamma\in \{-\eta,\eta\}$  can be calculated by,
\begin{align}
N_\Gamma= {\displaystyle \int_{\Gamma}\frac{dN}{d\eta} d\eta}.
\label{average1}
\end{align}
\section{RESULTS AND DISCUSSION}
\subsection{$\langle p_z^2\rangle_\Gamma $ and $N_\Gamma$ for Pb+Pb collisions  at $2.76A$ TeV at the LHC}
In section III, we have shown that the longitudinal momentum distribution of particles plays a crucial role in determining the effects of GMC on the longitudinal flow decorrelation.  How can one determine phase space volume for the global momentum conservation in relativistic heavy-ion collisions?  Ideally, the full phase-space volume should be considered.  
In reality,  some particles might reside at the high $\eta$ region of the pseudorapidity distribution (i.e., large longitudinal momentum) originating from the spectators far away from the collision zone.  Such particles should not be considered as they are not the products of the collision zone.  Therefore, we only consider a finite phase space region for global momentum conservation to mimic the real case.\\
\begin{figure}[htbp!]
	\centering
	{\includegraphics*[scale=0.40,clip=true]{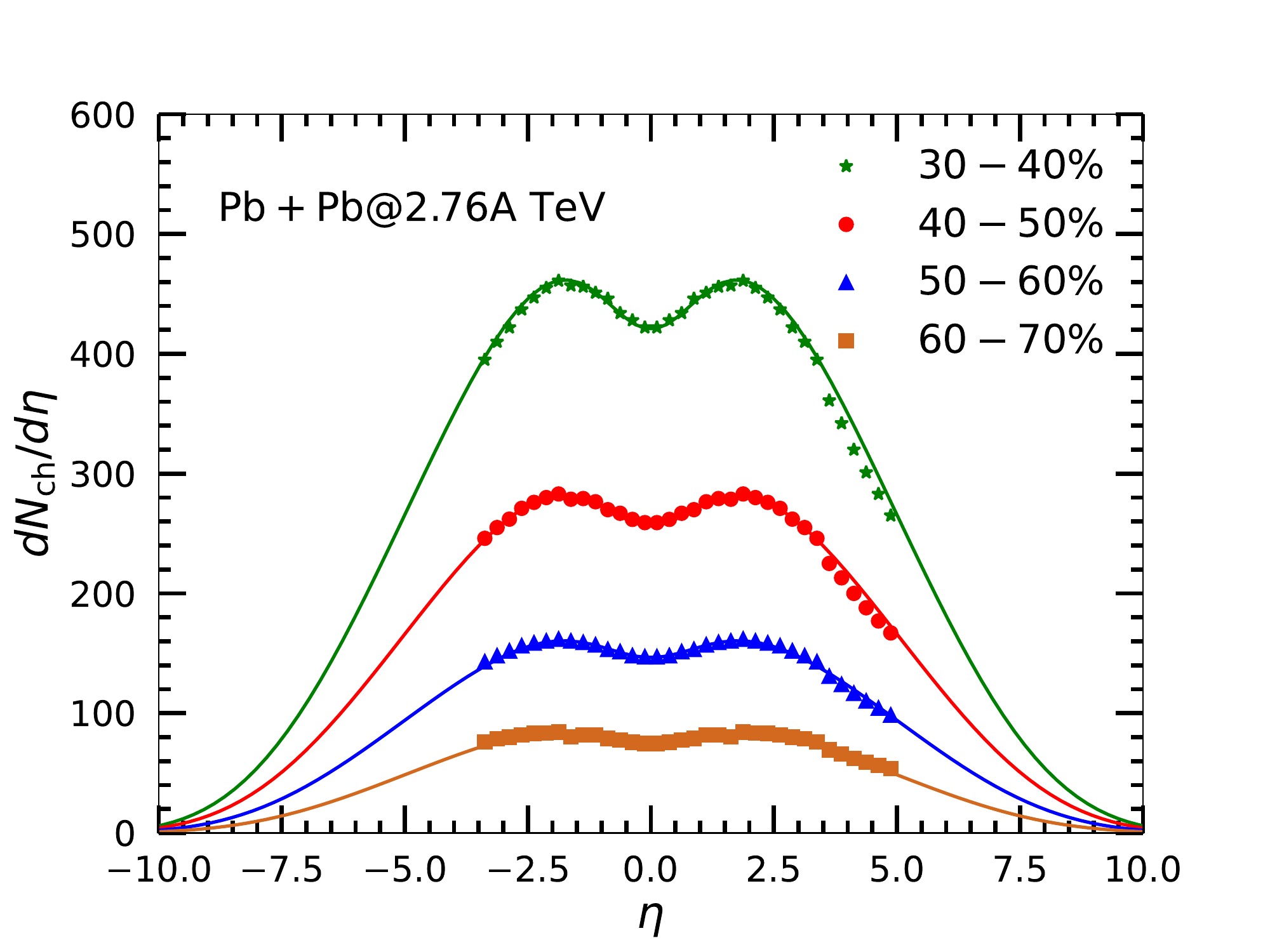}}
	\vspace{-3mm}
	\caption {(Color online)  Charged particle pseudorapidity density distributions for $30$-$40\%$,$40$-$50\%$, $50$-$60\%$ and $60$-$70\%$ centrality bins in Pb+Pb collisions at 2.76A TeV, where the charged particle pseudorapidity distributions for four centrality bins from ALICE are fitted with the parameters presented in TABLE~\ref{table:obs}. }
	\label{fig1}
\end{figure}
We use the relativistic hydrodynamical framework described in section IV to fit the experimental data of charged particle pseudorapidity distributions for four different centrality bins (i.e., $30$-$40\%$, $40$-$50\%$, $50$-$60\%$ and $60$-$70\%$ ) of Pb+Pb collisions at $2.76A$ TeV at ALICE~\cite{ALICE:2015bpk}.  Consequently, we evaluate various needed quantities to determine the effects of GMC on longitudinal decorrelation parameters $r_n$.

\begin{figure}[tb]
	\centering
	{\includegraphics*[scale=0.40,clip=true]{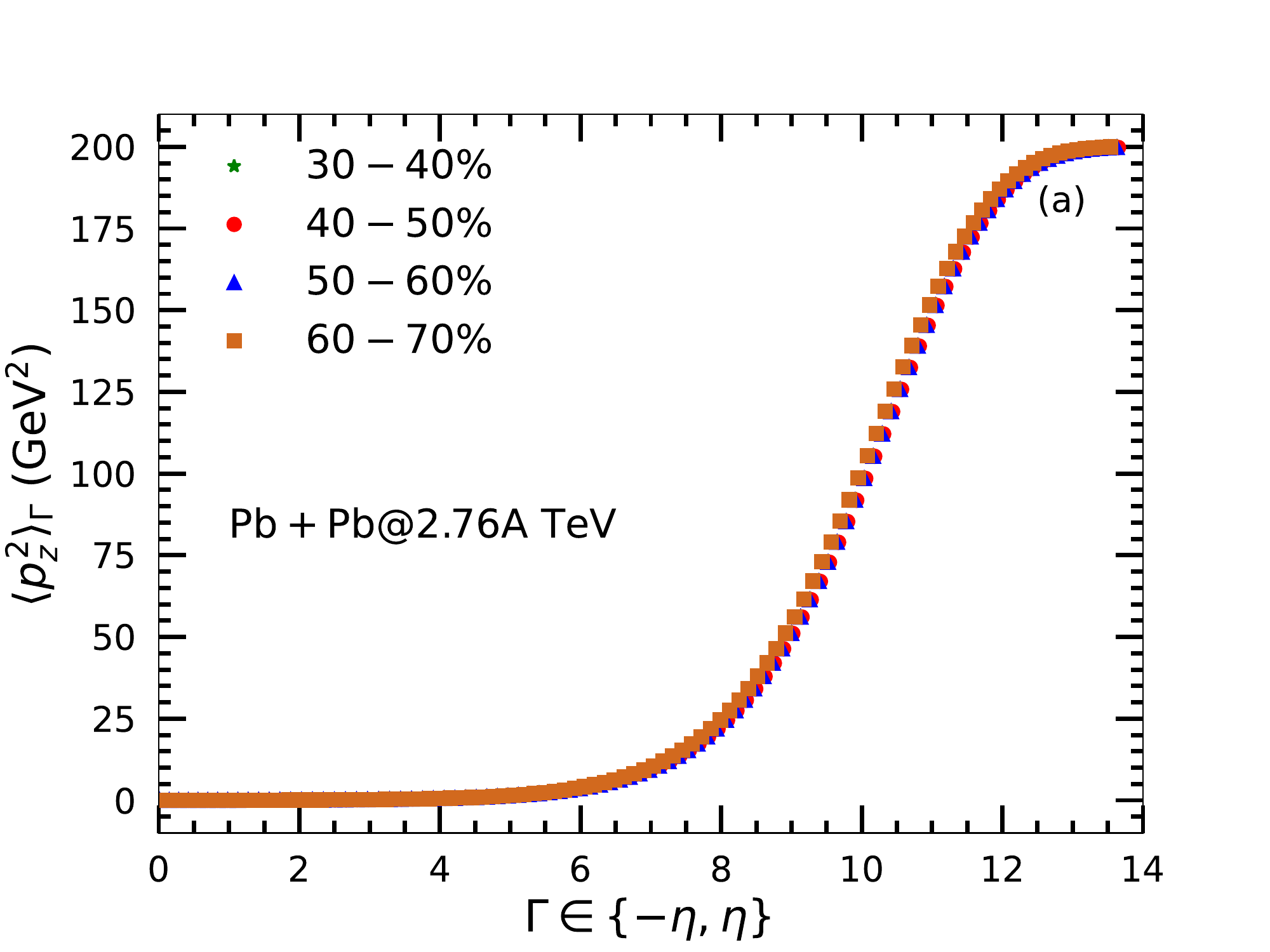}}
	{\includegraphics*[scale=0.40,clip=true]{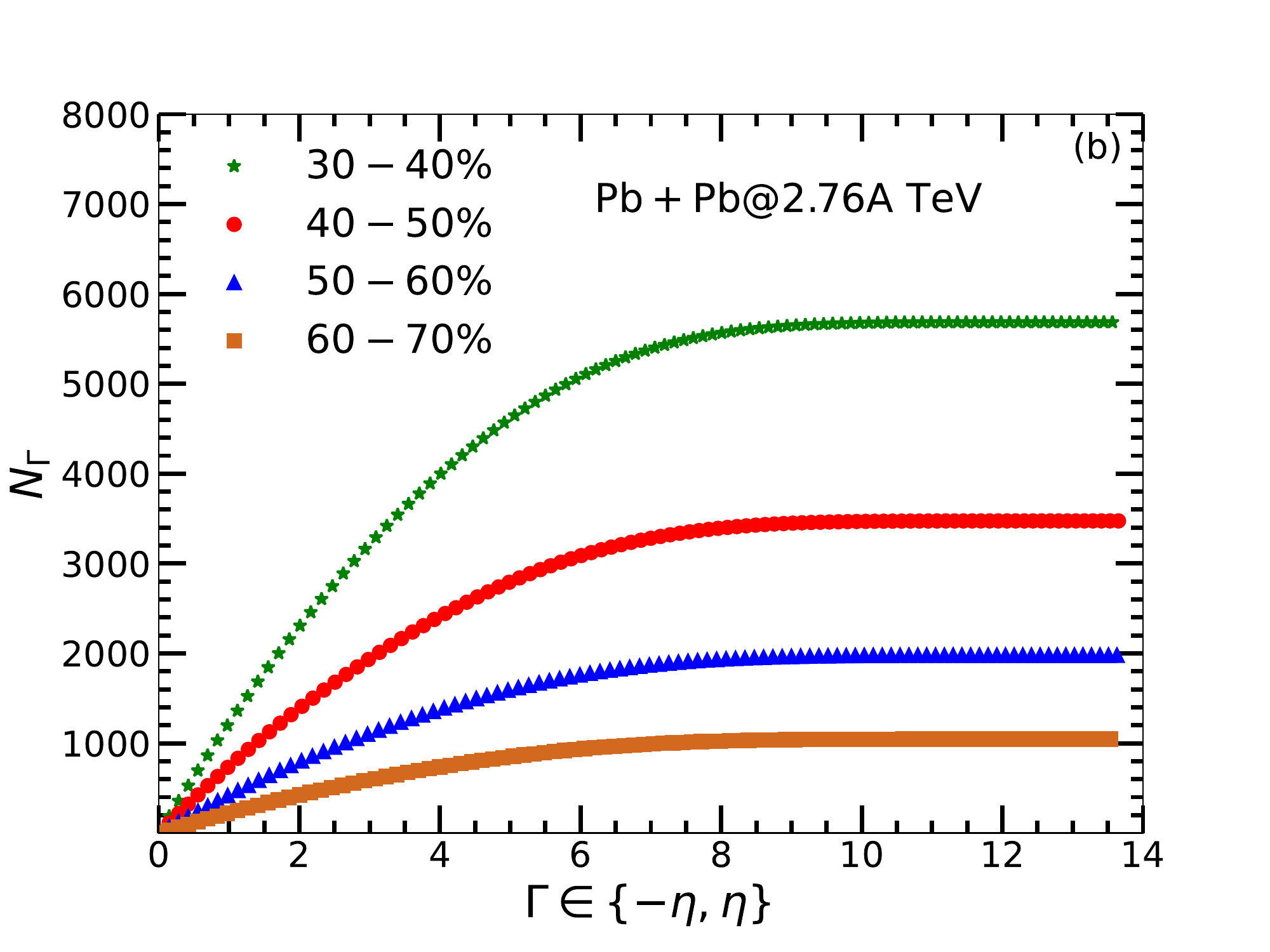}}
	\vspace{-3mm}
	\caption {(Color online) (a) $\langle p_z^2\rangle_\Gamma$  and (b) $ N_\Gamma$ as a function of phase space region $\Gamma\in \{-\eta,\eta\}$  for $30$-$40\%$,$40$-$50\%$, $50$-$60\%$ and $60$-$70\%$ centrality bins in Pb+Pb collisions at $2.76A$ TeV. }
	\label{fig2}
	
\end{figure}
We use the fit parameters shown in TABLE I (obtained from Ref.~\cite{Ze-Fang:2017ppe}) to reproduce the pseudorapidity distribution of charged hadrons from Pb+Pb collisions at $2.76A$ TeV. The average mass ($\bar{m}$) of all charged hadrons is taken as $0.24$ GeV, and the value of freeze-out temperature ($T_f$) is taken as $0.09$ GeV. The central pseudorapidity density ($dN_{\rm ch}/d\eta|_{\eta=0}$) for $30$-$40\%$, $40$-$50\%$, $50$-$60\%$ and $60$-$70\%$ centrality bins in Pb+Pb collisions at $2.76A$ TeV are taken as $422.0$, $259.1$, $147.1$, and $74.7$, respectively. As shown in Fig.~\ref{fig1}, the parameter set presented in TABLE-I can satisfactorily describe the pseudorapidity distributions of charged hadrons for all four centrality bins. We further approximate the total number of particles as 1.5 times the charged hadron multiplicity.

\begin{table*}[htbp]
\caption{Fit parameters in Fig.~\ref{fig1} using Eq.~(\ref{ydist}) for Pb+Pb collisions at 2.76A TeV~\cite{Ze-Fang:2017ppe}, where the auxiliary values of $T_f=0.09$ GeV and $\bar{m}=0.24$ GeV have been utilized.}
\label{table:obs}
\centering
\begin{tabular}{p{50pt}p{50pt}p{50pt}p{50pt}p{50pt}}
\hline
\hline
&$\frac{dN_{\rm ch}}{d\eta}\bigg|_{\eta=0}$ & $\lambda$ & $\sigma$& $T_{\rm eff}{\rm (GeV)}$  \\ 
\hline
		        $30$-$40\%$& 422.0 & 1.04 & 0.88 & 0.27\\
			\hline			
			$40$-$50\%$& 259.1 & 1.04 & 0.92 & 0.27\\
			\hline
			$50$-$60\%$& 147.1 & 1.04 & 0.91 & 0.27\\
			\hline
			$60$-$70\%$& 74.7 & 1.04 & 0.87 & 0.27\\
\hline
\hline
\end{tabular}
\end{table*}

In Fig.~\ref{fig2}, we show the $\langle p_z^2\rangle_\Gamma$ and $\langle N\rangle_\Gamma$ of all particles as a function of phase-space interval $\Gamma$ (i.e., pseudorapidity range) for Pb+Pb collisions at $2.76A$ TeV at the LHC.  
The $\langle p_z^2\rangle_\Gamma$ over a phase-space interval $\Gamma\in \{-\eta,\eta\}$ is computed by using Eq.~(\ref{average1}). In Fig.~\ref{fig2}(a), the $\langle p_z^2\rangle_\Gamma$ as function of $\Gamma$ for four centrality $30$-$40\%$, $40-50\%, 50-60\% $ and $60$-$70\%$ in Pb+Pb collisions at $2.76A$ TeV are presented. We find that the value of $\langle p_z^2\rangle_\Gamma$ first increases slowly up to $\Gamma\in \{-6,6\}$, then it starts to increase exponentially and finally saturates above $\Gamma\in \{-12,12\}$. The values of $\langle p_z^2\rangle_\Gamma$ for all four centrality bins are close to each other.  
In Fig.~\ref{fig2}(b), the $\langle N\rangle_\Gamma$ as a function of $\Gamma$  for the above centrality bins are shown. Considering a larger phase-space region, the average number of particles inside first increases sharply. Finally, it saturates above the pseudorapidity window $\Gamma\in \{-8,8\}$. We also see that the average number of particles for the $30$-$40\%$ centrality bin is almost  1.7, 3, and 6 times as large as $40$-$50\%$, $50$-$60\%$, and $60$-$70\%$ centrality bins, respectively.\\
\subsection{$R_2$ and $R_3$ for Pb+Pb collisions at $2.76A$ TeV  at the LHC}
\begin{figure*}[htbp!]
	\centering
	{\includegraphics*[scale=0.40,clip=true]{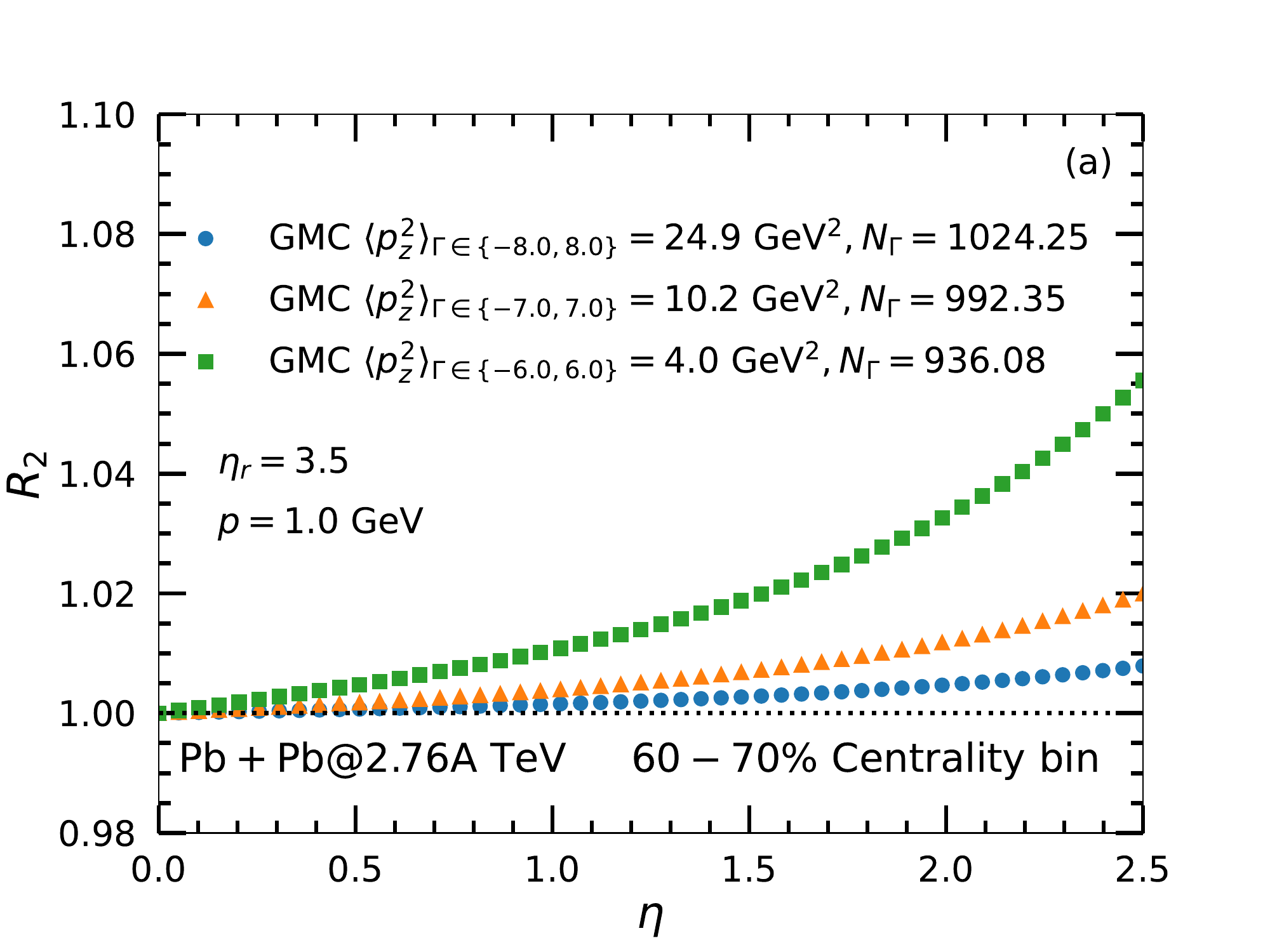}}\hspace{-0.1cm}
	{\includegraphics*[scale=0.40,clip=true]{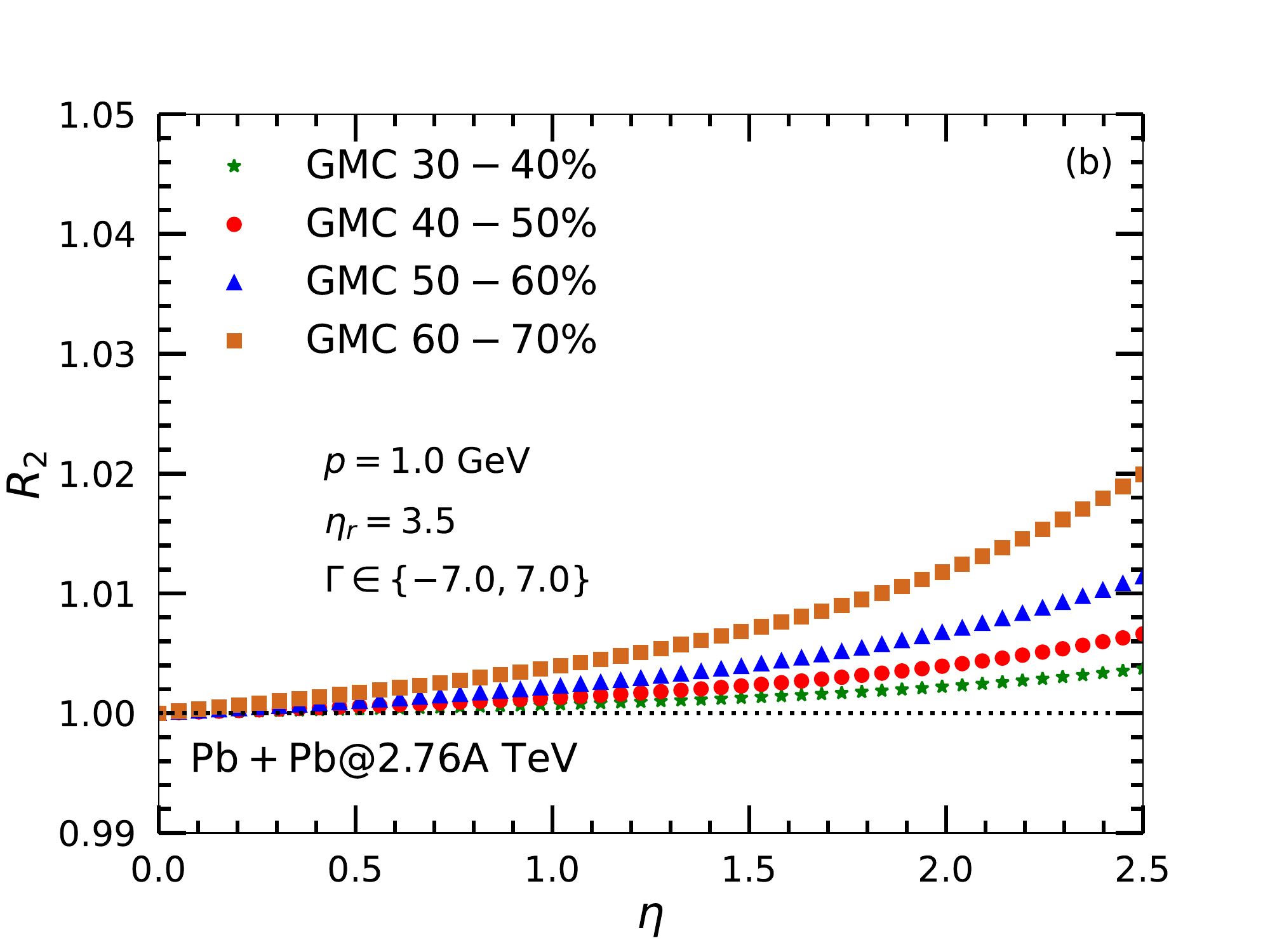}}\vspace{-0.1cm}
	{\includegraphics*[scale=0.40,clip=true]{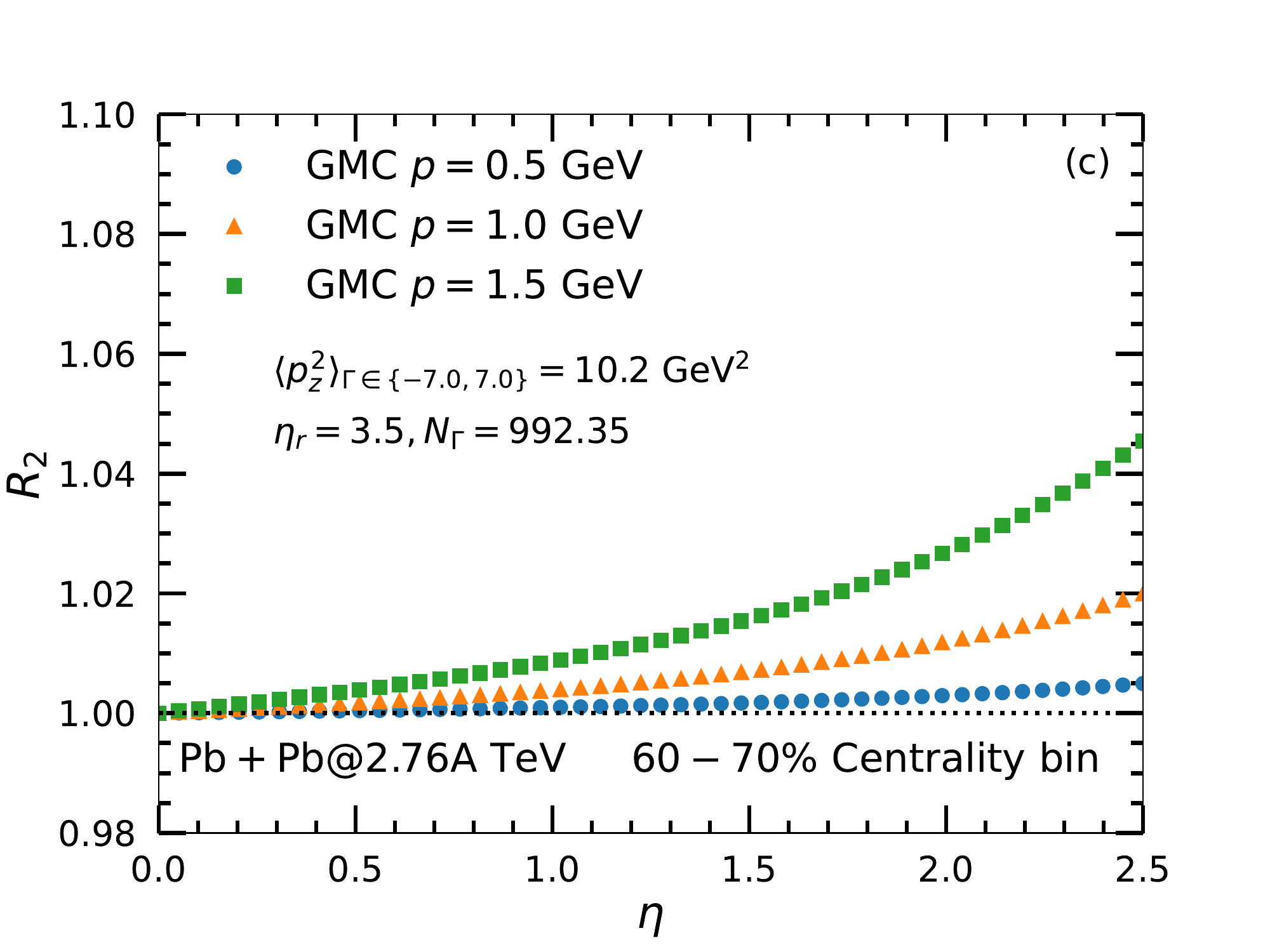}}\hspace{-0.1cm}
	{\includegraphics*[scale=0.40,clip=true]{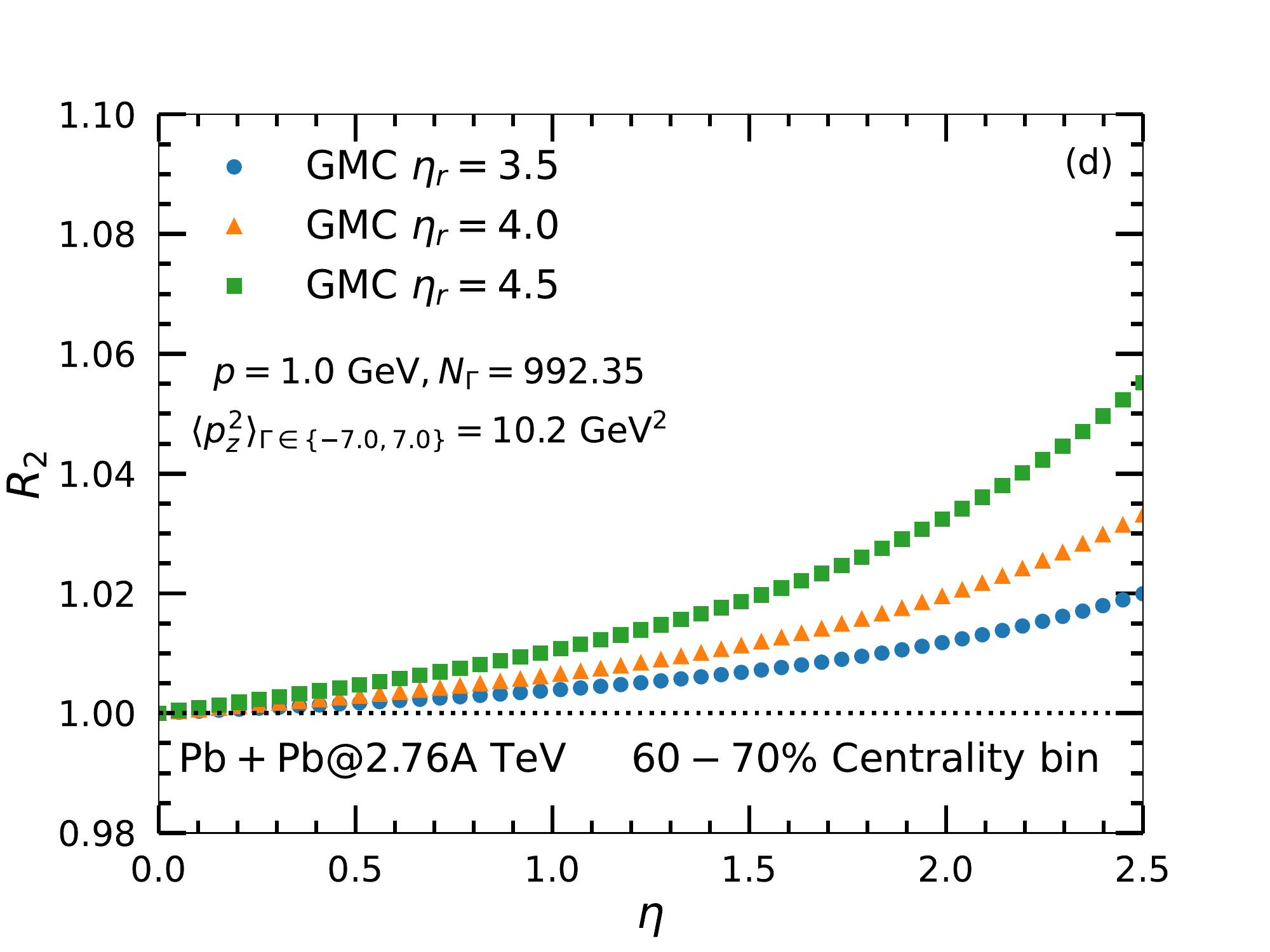}}
	\vspace{-3mm}
	\caption {(Color online)  The modification factor of longitudinal decorrelation $R_2$ as a function of $\eta$ for (a) different phase-space volumes with GMC constraint,  (b) different centrality bins, (c) different momenta, and (d) different reference pseudorapidities in Pb+Pb collisions at $2.76A$ TeV. }
	\label{fig3}
\end{figure*}

With the above information about $\langle N\rangle_\Gamma$ and $\langle p_z^2\rangle_\Gamma$,  we present the behavior of modification factor $R_2$ as a function of $\eta$  for Pb+Pb collisions at $2.76A$ TeV in Fig.~\ref{fig3}. In Fig.~\ref{fig3}(a), we show the $\eta$ dependence of $R_2$  for 60-70\% centrality bin for three selected phase volumes (i.e., $\Gamma\in\{-\eta,\eta\}$).  We see that $R_2$ is greater than one and increases with $\eta$ due to the constraint of the GMC effect. Among three chosen phase space regions, i.e, $\Gamma\in\{-8,8\},\{-7,7\}$, and $\{-6,6\}$, the effects on $R_2$ are found to be largest for the $\Gamma\in\{-6,6\}$.  It indicates that the effect of GMC on $R_2$ tends to increase if the phase space volume with the GMC constraint gets smaller. These features can be understood according to the modification factor $R_2$ shown in Eq.~(\ref{rnwogmc3}).  Note that the recent ALICE result supports that the baryon number follows a global baryon number conservation in Pb+Pb collisions at the LHC~\cite{ALICE:2019nbs}.

In Fig.~\ref{fig3}(b), we show $R_2$ as a function of $\eta$ for various centrality bins for Pb+Pb collisions at $2.76A$ TeV, where we assume that the GMC constraint is valid within $\Gamma\in\{-7,7\}$ for all the centrality bins.  We see that GMC impacts the peripheral collisions more than the central collisions due to the smaller number of particles $N$ in peripheral collisions.
\begin{figure*}[htbp!]
	\centering
	{\includegraphics*[scale=0.40,clip=true]{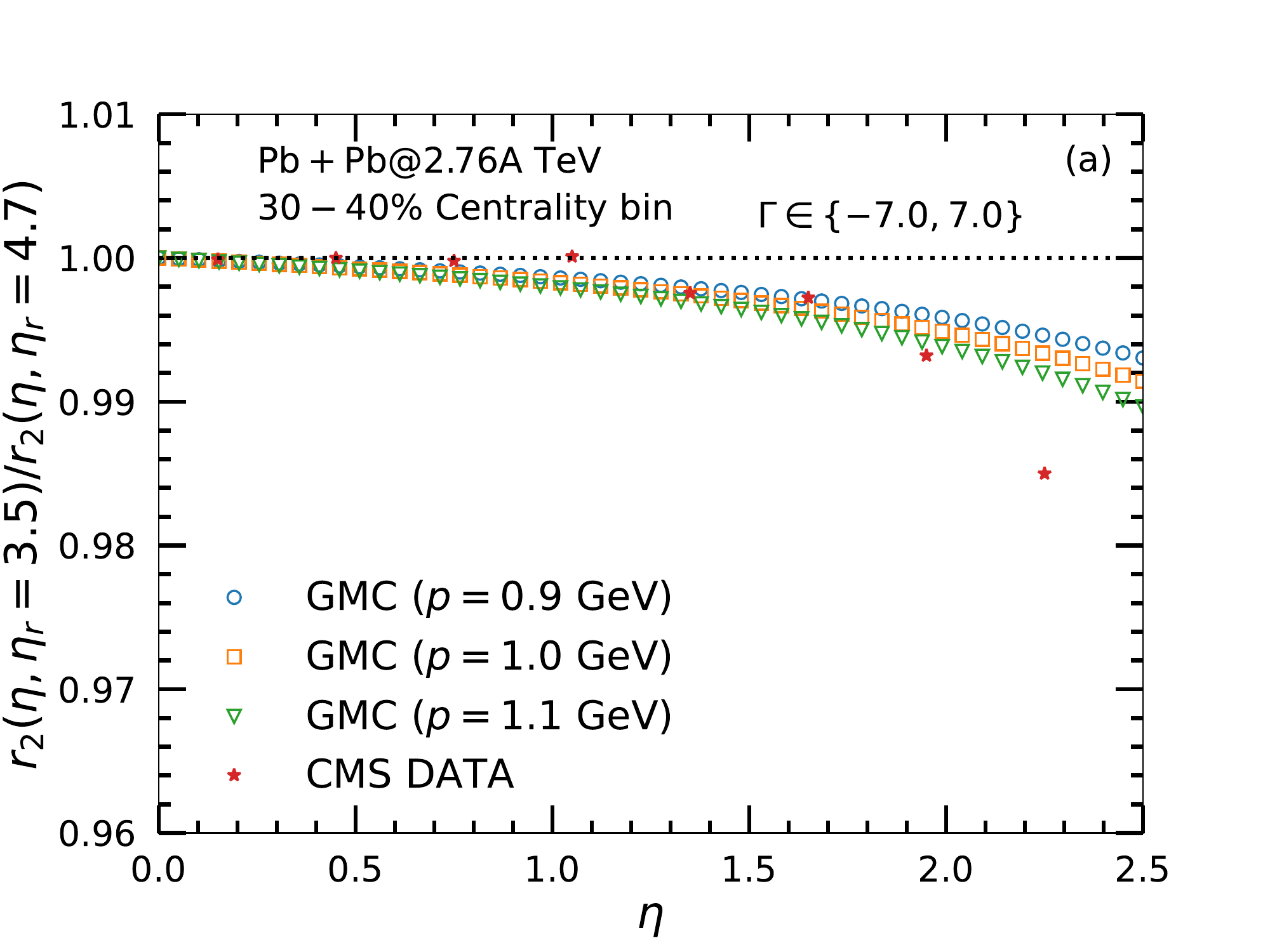}}\hspace{-0.1cm}
	{\includegraphics*[scale=0.40,clip=true]{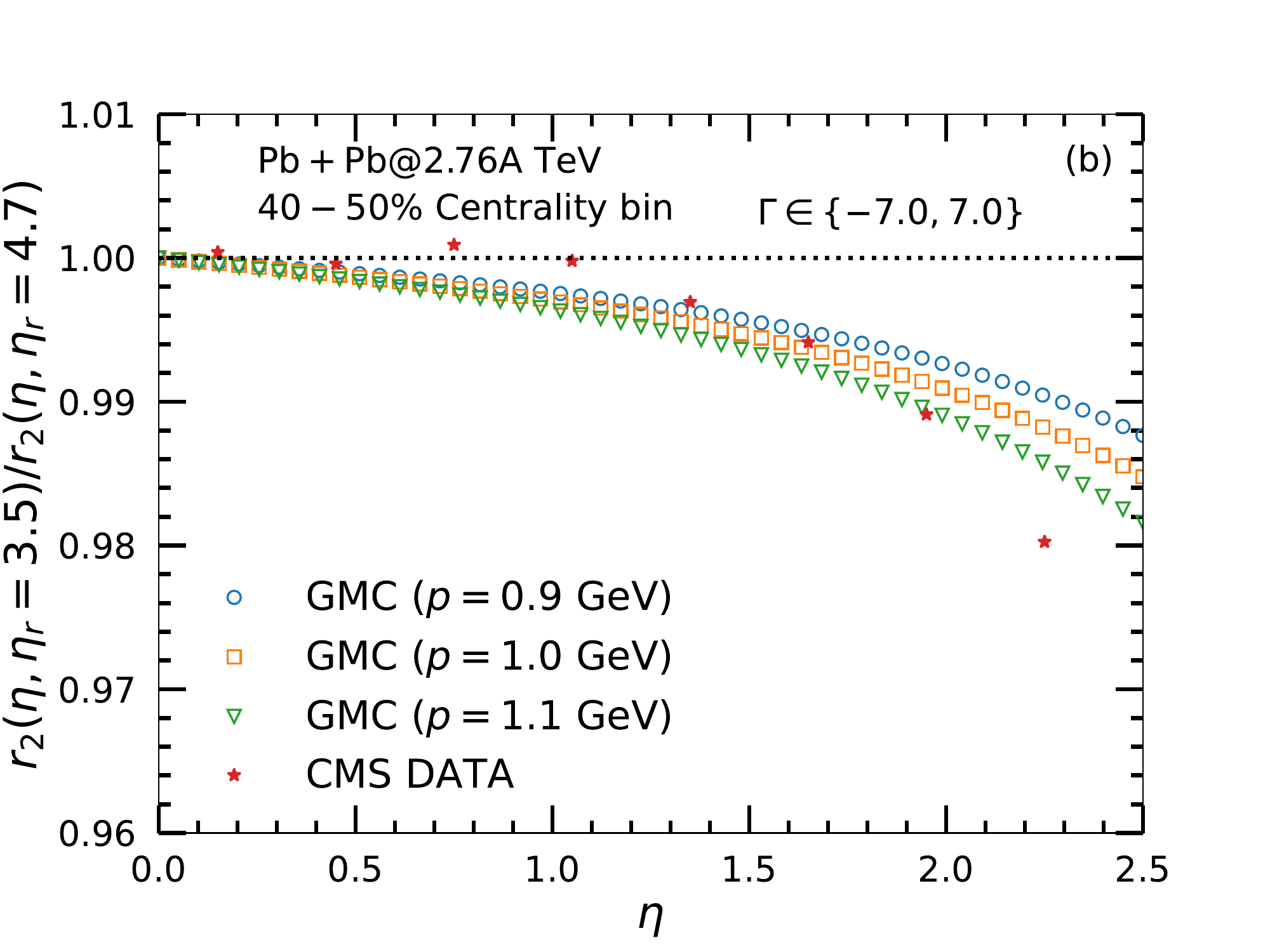}}\hspace{-0.1cm}
		{\includegraphics*[scale=0.40,clip=true]{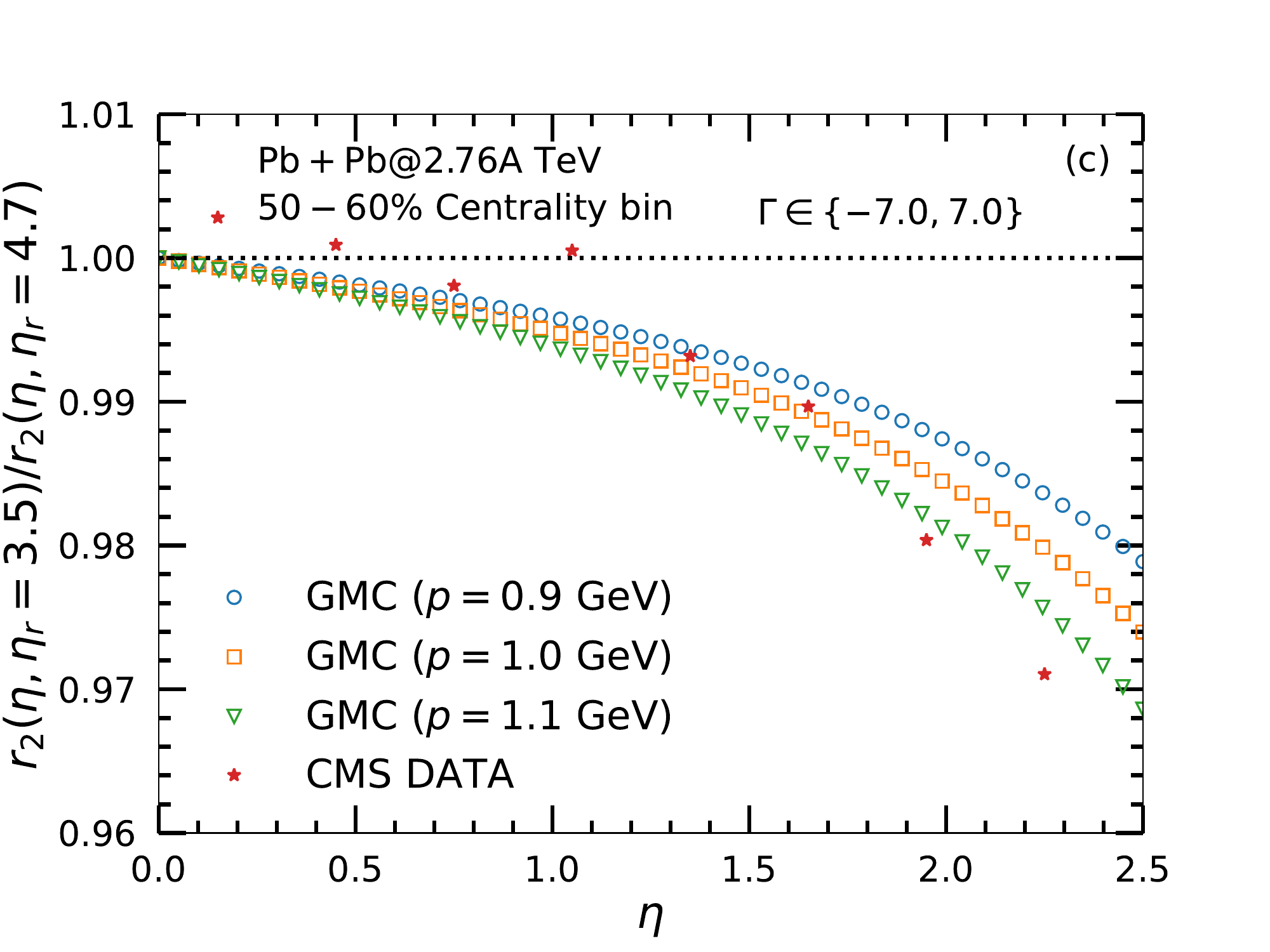}}\hspace{-0.1cm}
	\vspace{-3mm}
	\caption {(Color online)  The ratio of $r_2(\eta_1,\eta_r=3.5)$ over  $r_2(\eta_1,\eta_r=4.7)$  as a function of $\eta$ for (a) $30$-$40\%$, (b) $40$-$50\%$ and (c) $50$-$60\%$ centrality bins in Pb+Pb collisions at $2.76A$ TeV. }
	\label{fig4}
\end{figure*}
In Fig.~\ref{fig3}(c), we show the $\eta$ dependence of $R_2$ for the different transverse momenta $p$ of particles for the centrality bin of 60-70\% in Pb+Pb collisions at $2.76A$ TeV, where we choose $p_1=p_2=p_3=p$. For $p$ = 1.5 GeV, the GMC effect is the largest compared to the same obtained from the smaller momentum values. It indicates that the effect of GMC increases with the transverse momenta of particles. It should be pointed out that so far, we have computed the momentum-dependent observable with fixed momentum values; however, in experiments, the longitudinal decorrelation was measured for all charged particles within the momentum range, e.g., $0.3$ GeV $<p<3.0$ GeV. We expect that the behavior of the momentum-integrated decorrelation under the influence of GMC is similar to that shown here (see the Appendix please).  

In Fig.~\ref{fig3}(d), we show the $\eta$ dependence of $R_2$ for different reference pseudorapidity  ($\eta_r$) for the centrality bin of 60-70\% in Pb+Pb collisions at $2.76A$ TeV. We find that the effect of GMC increases as $\eta_r$ gets further away, resulting in a smaller longitudinal decorrelation in the larger $\eta_r$ case. A similar reference pseudorapidity dependence of $r_n(\eta,\eta_r)$ has been observed for Pb+Pb  and $p+$Pb collisions at the LHC, where the decorrelation effect is weaker if a larger reference pseudorapidity bin is applied~\cite{CMS:2015xmx}.

In Fig.~\ref{fig4}, we further elaborate on the reference pseudorapidity dependence of the GMC-induced results by comparing the ratio of two longitudinal decorrelation coefficients at separate reference pseudorapidity bins [$r_2(\eta,\eta_{rA})/r_2(\eta,\eta_{rB})$] with the CMS data~\cite{CMS:2015xmx}. Here we calculate the ratio of $r_2$ with different $\eta_r$ for different centrality bins in Pb+Pb collisions and compare our findings with the results obtained from the experimental measurements. In this way, we hope to cancel the flow contribution to show a more evident effect of GMC.  Surprisingly, by invoking the effect of GMC, we can describe the reference pseudorapidity dependence of the data. In Fig.~\ref{fig4} (a), (b) and (c), we compare the experimental ratios of $r_2(\eta,\eta_r=3.5)/r_2(\eta,\eta_r=4.7)$ with our theoretical results for $30$-$40\%$, $40$-$50\%$ and $50$-$60\%$ centrality bins in Pb+Pb collisions at $2.76A$ TeV,  respectively. Our results with momentum $p\sim 1.0\pm 0.1$ GeV can satisfactorily describe the data for all three centrality bins. We also find that such reference pseudorapidity dependence is even enhanced if GMC is considered for a smaller region of phase space. 
On the other hand, Figs.~\ref{fig4} (a)-(c) show that the GMC effect on the reference pseudorapidity dependence is sensitive to the transverse momentum $p$. The higher the transverse momentum, the further the ratio is from one. We notice that the reference pseudorapidity dependence of the longitudinal flow decorrelation can be observed in a hydrodynamic model if the initial flow is included, which has been argued to be caused by the nonthermalized mini-jets in the initial state~\cite{Pang:2018zzo}. Since the slope of the transverse momentum spectra is increased by the initial flow~\cite{Pang:2012he}, our results are consistent with the hydrodynamic finding in fact. Therefore, we suggest studying a more significant GMC effect on the longitudinal flow decorrelation at higher transverse momentum. \\

\begin{figure*}
	\centering
	{\includegraphics*[scale=0.40,clip=true]{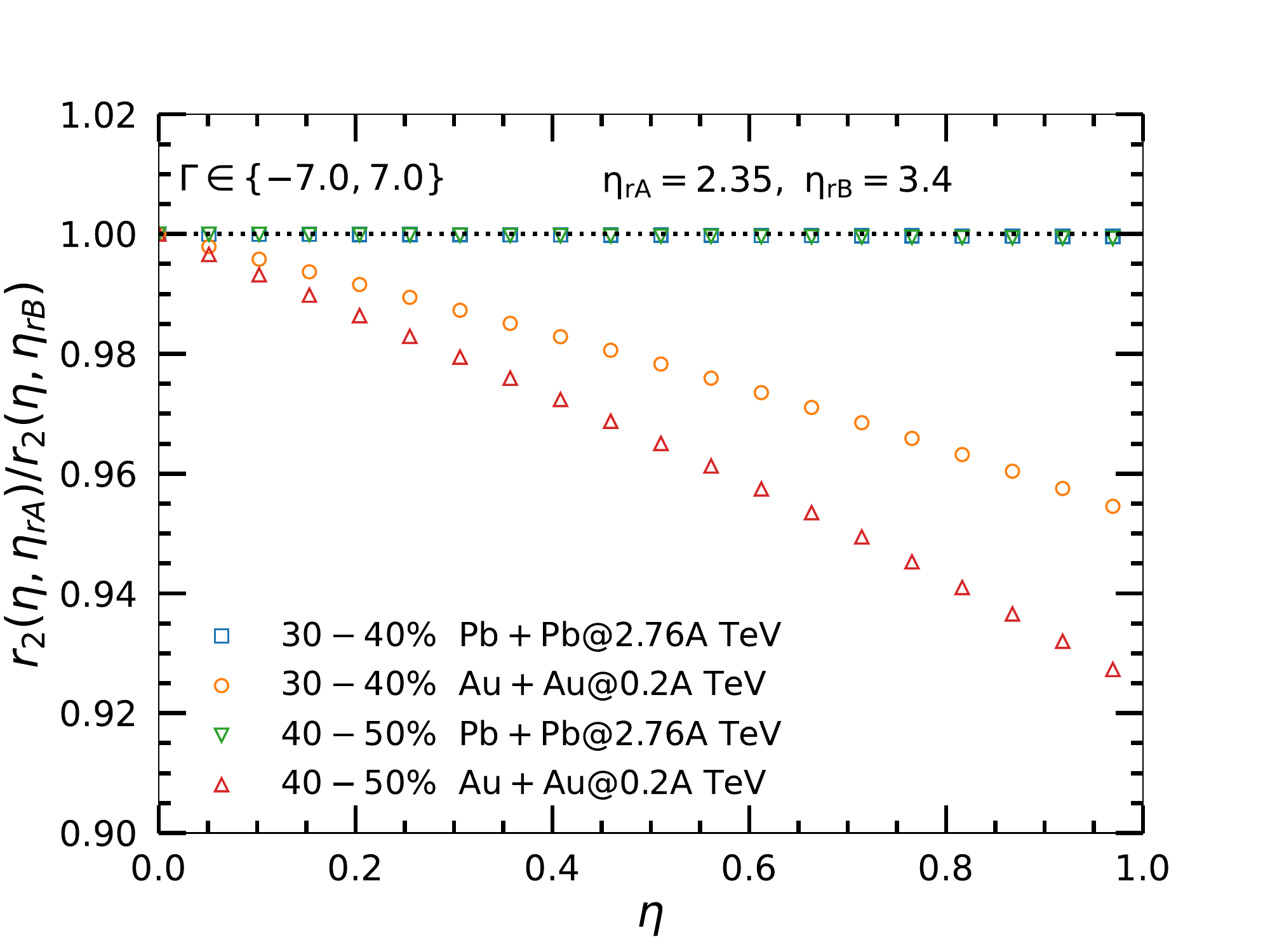}}
	\vspace{-3mm}
	\caption {(Color online)  The ratio of $r_2(\eta_1,\eta_r=2.35)$ over  $r_2(\eta_1,\eta_r=3.4)$ as a function of $\eta$ for $30$-$40\%$ and $40$-$50\%$ Pb+Pb collisions at $2.76A$ TeV and Au+Au collisions at $0.2A$ TeV. }
\label{fig5}
\end{figure*}

The pseudorapidity dependence of  $r_2(\eta,\eta_{rA})/r_2(\eta,\eta_{rB})$ ratio is further explored for Au+Au collisions at the RHIC energy in Fig.~\ref{fig5}.  We simultaneously compare the ratios of $r_2(\eta,\eta_r=2.35)/r_2(\eta,\eta_r=3.4)$ for 30-40\% and 40-50\% centrality bins in Pb+Pb collisions at $2.76A$ TeV and Au+Au collisions at $0.2A$ TeV.  We find that the ratio is more significantly affected in Au+Au collisions at the RHIC energy than in Pb+Pb collisions at the LHC energy. We also observe a larger ratio for more peripheral collisions at RHIC. Note that the two results for two centrality bins in Pb+Pb collisions at $2.76A$~TeV are almost overlapping and consistent with one for this $\eta_r$ selection. We notice that the STAR collaboration is analyzing the longitudinal correlation coefficients in isobar collisions at $0.2A$ TeV. It is a good potential collision system to verify the characteristics of the GMC effect since the multiplicity in isobar collisions is much smaller than in Au+Au collisions.

In addition, we follow Eq.~(\ref{rnwogmc4}) to explore the behavior of $R_3$ in Pb+Pb collisions at $2.76A$ TeV. We find that $R_3$ behaves similarly to $R_2$ because the dominant contribution comes from the terms which consist of $v_3$ only (if with the $v_3$ only terms, $R_3\equiv R_2$). We further check the contribution of  $v_2$ terms in Eq.~(\ref{rnwogmc4}).  By assuming $\langle \cos[n(\psi_n(\eta_a) -\psi_n(\eta_r))]\rangle \approx e^{-F_n |\eta_a - \eta_r |}$ (for $n$=2 and 3) and the values of $F_2$, $F_3$, $v_2$ and $v_3$ to be $0.02$, $0.04$, $0.05$ and $0.025$, respectively, we find that the inclusion of $v_2$-involving terms introduces a reduction in $R_3$ by $10-15\%$.

We would like to point out that our employed parametric form of the rapidity distribution is for an ideal fluid system. However, a small specific shear viscosity has been extracted from the experimental data collected in heavy-ion collision experiments at RHIC and LHC~\cite{Heinz:2013th,Gale:2013da}. In general, the longitudinal fluid evolution is decelerated due to the shear viscosity~\cite{Teaney:2003kp,Bozek:2007qt}, which may cause a decrease in $\langle p_z^2\rangle_\Gamma$. On the other hand, the shear viscosity can lead to higher transverse velocities (radial flow), which increases the transverse momenta of particles~\cite{Romatschke:2007mq}. Since both factors help enhance the GMC effect, we expect that the GMC effect on the longitudinal flow decorrelation increases when the shear viscosity effect is taken into account. 

\section{Summary}

In summary, we explore the effect of the global momentum conservation on the longitudinal flow decorrelation coefficients. We first present the analytic forms of two-particle azimuthal cumulants and the decorrelation coefficients in the presence of hydro-like flow and the GMC constraint. Subsequently, we explore the modification effect of GMC on the longitudinal flow decorrelations in Pb+Pb collisions at the LHC energy and Au+Au collisions at the RHIC energy. The modification factors $R_2$ and $R_3$ of the second-order and third-order longitudinal flow decorrelation coefficients weaken the longitudinal flow decorrelation due to the presence of GMC. We find that the modification factors are sensitive to the total number of involved particles ($N$), the average longitudinal momentum ($\langle {p_z^2}\rangle_F$), the transverse momentum ($p$), and the reference pseudorapidity ($\eta_r$). Our results of $r_2(\eta,\eta_{rA})/r_2(\eta,\eta_{rB})$ ratios are consistent with the experimental measurements. We also predict that the modification effect is stronger for RHIC than for the LHC. Our finding suggests that the effect of GMC constraint should be taken into account when studying the longitudinal flow decorrelation in relativistic heavy-ion collisions.

\begin{acknowledgments}
This work is supported by the National Natural Science Foundation of China under Grants No. 12150410303, No.12147101, No. 11890714, No. 11835002, No. 11961131011, No. 11421505, the National Key Research and Development Program of China under Grant No. 2022YFA1604900, the Strategic Priority Research Program of Chinese Academy of Sciences under Grant No. XDB34030000, and the Guangdong Major Project of Basic and Applied Basic Research under Grant No. 2020B0301030008.  
\end{acknowledgments}

\section*{Appendix}

\begin{figure}[htbp!]
	\centering
	{\includegraphics*[scale=0.40,clip=true]{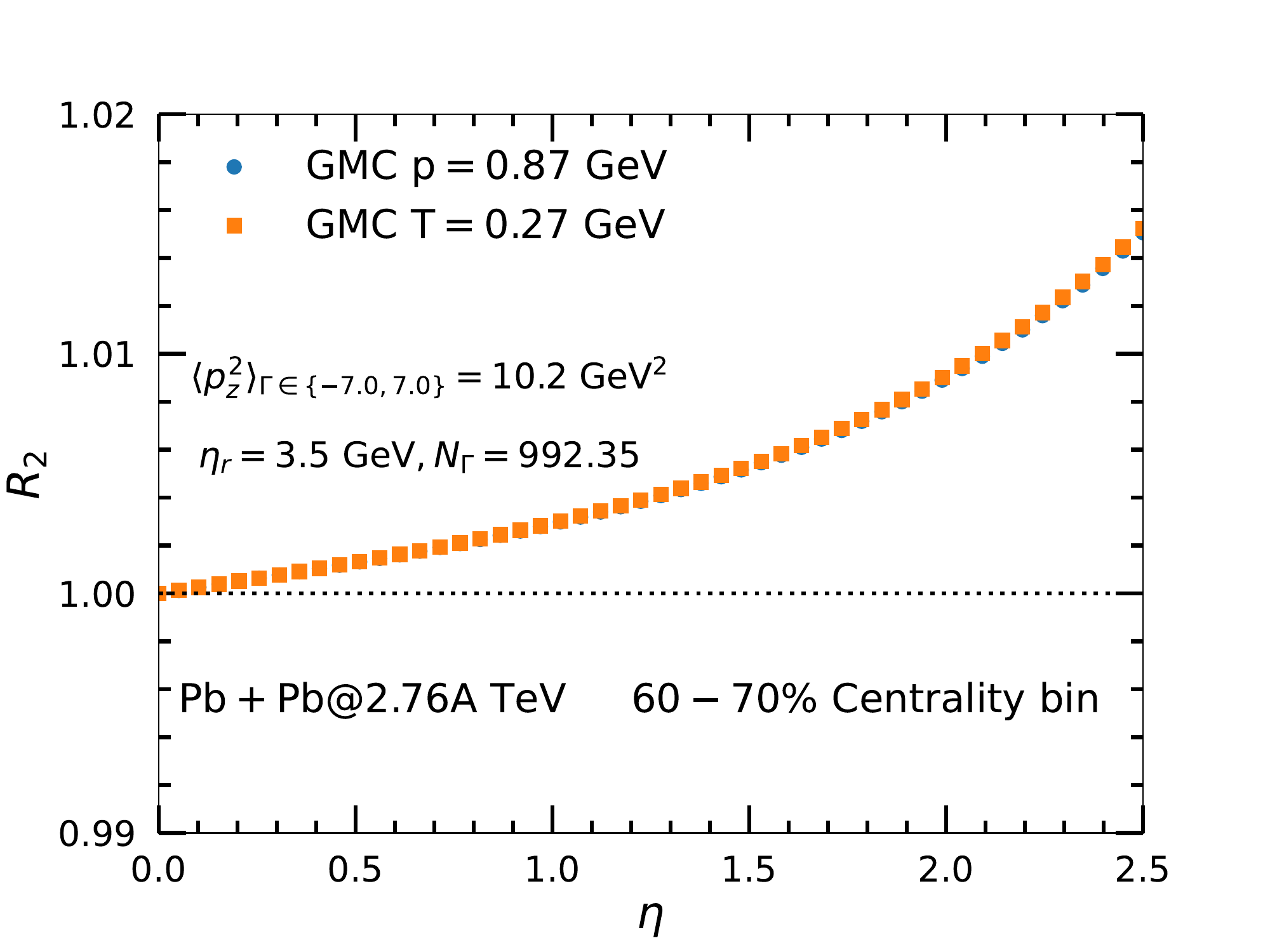}}
	\caption{ (Color online) The modification factor $R_2$ corresponding to the integrated longitudinal decorrelation coefficient for $60$-$70\%$ Pb+Pb collisions at 2.76A TeV at the LHC, in comparison with the differential $R_2$ corresponding to $p\approx0.87$ GeV.}
	\vspace{-3mm}
	\label{fig7}
	
\end{figure}

In this appendix, we present the integrated longitudinal decorrelation coefficient, i.e., $r_2(\eta,\eta_r)$, under the influence of GMC. $F(\bf{p})$ is the single-particle distribution function having a form as shown in Eq.~(\ref{hydrolike}). Considering the A+A collision system to be symmetric, we further assume the following  forms of the pseudorapidity and momentum-dependent function $g(p,\eta)$ and $v_2(p,\eta)$~\cite{Bzdak:2010fd}:
\begin{eqnarray}
 &g(p,\eta)=\frac{1}{T^2} \exp\big(-\frac{p}{T}\big) \ h_1(\eta),
 \ \ \ v_2(p,\eta)=\ p \  h_2(\eta)
\end{eqnarray}
where, $h_1(\eta)$ and $h_2(\eta)$ are two even functions of pseudorapidity. Now,
introducing the above forms of the single and joint probability distribution functions into  Eq.~(\ref{rndef}), and performing a momentum $p$ integration over $p_l$ to $p_h$, we finally obtain the integrated form of $r_2(\eta,\eta_r)$ :
\begin{eqnarray}
 & r_2(\eta,\eta_r)|^{\rm \sf GMC+Flow}=r_2(\eta,\eta_r)|^{\rm \sf Flow} \  \ \times  \ \ R_2, \nonumber\\
	& R_2\approx\bigg[\frac{N \langle p_z^2\rangle_F \left(x(p_h,T) e^{p_l/T}-e^{p_h/T} x(p_l,T)\right)^2+\sinh ({\eta}) \sinh ({\eta_r}) \left(y(p_h,T) e^{p_l/T}-e^{p_h/T} y(p_l,T)\right)^2}{N \langle p_z^2\rangle_F \left(x(p_h,T) e^{p_l/T}-e^{p_h/T}x(p_l,T)\right)^2-\sinh ({\eta}) \sinh ({\eta_r}) \left(y(p_h,T) e^{p_l/T}-e^{p_h/T} y(p_l,T) \right)^2}\bigg]\\
& {\rm where, \ \ \ \ }	x(p,T)=p^2+2 p T+2 T^2 { \ \ \ \ \ \rm and \ \ \ }  y(p,T)=p^3+3 p^2 T+6 p T^2+6 T^3	\nonumber\
\end{eqnarray}

In Fig.~\ref{fig7}, we show the $\eta$ dependence of $R_2$ corresponding to the transverse momentum integrated longitudinal decorrelation coefficient for $60$-$70\%$ Pb+Pb collisions at 2.76A TeV at the LHC.  The integration limit has been chosen as $0.3$ GeV $<p<3.0$ GeV~\cite{CMS:2015xmx}. Subsequently, we compare our result with the differential $R_2$ obtained using Eq.~(\ref{rnwogmc3}) for the same collision system, and we observe that the result for $p\approx0.87$ GeV/ agrees with the integrated longitudinal decorrelation coefficient. This further proves that the above results at a mean $p$ value are reasonable to predict the behaviors of the integrated longitudinal decorrelation coefficients.

\end{CJK*}

\end{document}